\documentclass[%
 aip,
 jmp,%
 amsmath,amssymb,
 preprint,%
 fleqn,
]{revtex4-1}
\usepackage{graphicx}
\usepackage{dcolumn}
\usepackage{bm}
\begin{document}

\title{An analytical study on the existence of solitary wave and double layer solution of the well-known energy integral at $M = M_{c}$} 
\author{Animesh Das}
\author{Anup Bandyopadhyay}
\affiliation{ Department of Mathematics, Jadavpur
University,Kolkata - 700 032, India.}
\begin{abstract}
A general theory for the existence of solitary wave and double layer at $M=M_{c}$ has been discussed, where $M_{c}$ is the lower bound of the Mach number $M$, i.e., solitary wave and/or double layer solutions of the well-known energy integral start to exist for $M>M_{c}$. Ten important theorems have been proved to confirm the existence of solitary wave and double layer at $M=M_{c}$. If $V(\phi)(\equiv V(M,\phi))$ denotes the Sagdeev potential with $\phi$ is the perturbed field or perturbed dependent variable associated with the specific problem, $V(M,\phi)$ is well defined as a real number for all $M \in \mathcal{M}$ and for all $\phi \in \Phi$, and $V(M,0)=V'(M,0)=V''(M_{c},0)=0$, $V'''(M_{c},0)<0$ ($V'''(M_{c},0)>0$), $\partial V/\partial M < 0$ for all $M (\in \mathcal{M})>0$ and for all $\phi (\in \Phi)>0$ ($\phi (\in \Phi) <0$), where `` $' \equiv \partial/\partial \phi~$", the main analytical results for the existence of solitary wave and double layer solution of the energy integral at $M=M_{c}$ are as follows. \textbf{Result-1:} If there exists at least one value $M_{0}$ of $M$ such that the system supports positive (negative) potential solitary waves for all  $M_{c}<M< M_{0}$, then there exist either a positive (negative) potential solitary wave or a positive (negative) potential double layer at $M=M_{c}$. \textbf{Result-2:} If the system supports only negative (positive) potential solitary waves for $M>M_{c}$, then there does not exist positive (negative) potential solitary wave at $M=M_{c}$. \textbf{Result-3:} It is not possible to have coexistence of both positive and negative potential solitary structures (including double layers) at $M=M_{c}$. Apart from the conditions of \textbf{Result-1}, the double layer solution at $M=M_{c}$ is possible only when there exists a double layer solution in any right neighborhood of $M_{c}$. Finally, these analytical results have been applied to a specific problem on dust acoustic waves in nonthermal plasma in search of new results.
\end{abstract}
\maketitle 

\section{\label{sec:intro}Introduction}
The Sagdeev potential approach \cite{Sagdeev66} has been considered to investigate the existence of solitary wave and double layer at $M=M_{c}$, where $M_{c}$ is the lower bound of the Mach number $M$, i.e., solitary wave and/or double layer solutions of the energy integral start to exist for $M>M_{c}$. In most of the earlier works, \cite{Shukla78,Yu80,Bharuthram86,Baboolal88,Baboolal89,Baboolal90,Baboolal91,Cairns95a,Cairns95b,Cairns95c,Popel96,Mamun96a,Mamun96b,Xie98,Mendoza00,Gill03,Maharaj04,Maharaj06,Hellberg06,Verheest08,Baluku08,Tanjia08,Verheest09,Djebli09,das09,das10,Verheest10a} solitary wave and/or double layer solutions have been investigated for $M>M_{c}$. However, some recent investigations \cite{Verheest10b,Baluku10,Verheest10c,Baluku10a} have shown that finite amplitude solitary wave can exist at $M=M_{c}$ in the parameter regime where solitons of both polarities exist. The numerical observations \cite{Verheest10b,Baluku10,Verheest10c,Baluku10a} of the solitary wave solution of the energy integral at $M=M_{c}$, motivate us to set a general analytical theory for the existence of solitary wave and double layer solution of the energy integral at $M = M_{c}$. The present paper is purely analytical study on the existence of solitary wave and double layer solution of the energy integral at $M = M_{c}$. Without this analytical study, it is difficult to predict the existence of double layer solution at $M = M_{c}$. The analytical study also give the point in the parameter regime where the double layer solution at $M = M_{c}$ exists. So, we consider the following well known energy integral, which has been commonly used in various plasma environments. \cite{Shukla78,Yu80,Bharuthram86,Baboolal88,Baboolal89,Baboolal91,Cairns95a,Cairns95b,Cairns95c,Popel96,Mamun96a,Mamun96b,Xie98,Mendoza00,Gill03,Maharaj04,Maharaj06,Hellberg06,Verheest08,Baluku08,Tanjia08,Verheest09,Djebli09,das09,das10,Verheest10a,Verheest10b,Baluku10,Verheest10c,Baluku10a}
\begin{eqnarray}\label{energy int}
\frac{1}{2} \bigg(\frac{d\phi}{d\xi}\bigg)^{2}+V(\phi) = 0,
\end{eqnarray}
with $V(\phi) \equiv V(M,\phi)$ is the Sagdeev potential. Here, $V(M,\phi)$ is same as $V(\phi)$, i.e., the Mach number $M$ is omitted from the notation $V(M,\phi)$ when no particular emphasis is put upon it and let $\mathcal{M} \times \Phi$ is the domain of definition of the function $V(M,\phi)$, i.e., $V(M,\phi)$ is well defined as a real number for all $M \in \mathcal{M}$ and for all $\phi \in \Phi$. Hereafter, by the phrases `` for all $\phi > 0$ '', `` for all $\phi < 0$ '', and `` for all $M > 0$ '', we mean, respectively, `` all those $\phi \in \Phi$ which are strictly positive '', `` all those $\phi \in \Phi$ which are strictly negative '', and `` all those $M \in \mathcal{M}$ which are strictly positive ''.      

If `` $'$ " indicates differentiation of $V(M,\phi)$ with respect to $\phi$, then with the help of some definitions and some well-known theorems of real analysis, we have proved the following two theorems, which are necessary to prove the main theorems regarding the existence of solitary wave and double layer at $M=M_{c}$.\\
\textbf{Theorem 1 :}  If $V'''(M_{c},0)\neq 0$, then there exists a strictly positive real number $\phi_{\epsilon}$ ($>0$) such that $V'''(M_{c},0) V'''(M_{c},\phi) > 0$ for all $-\phi_{\epsilon} < \phi < \phi_{\epsilon}$.\\
\textbf{Theorem 2 :} If $V''(M_{c},0) = 0$ and $V'''(M_{c},0)\neq 0$, then there exists a strictly positive real number $\phi_{\epsilon}$ ($>0$) such that $V'''(M_{c},0) V''(M_{c},\phi) > 0$ for all $0 < \phi < \phi_{\epsilon}$ and $V'''(M_{c},0) V''(M_{c},\phi) < 0$ for all $-\phi_{\epsilon} < \phi < 0$.

If $V(M,0)=V'(M,0)=V''(M_{c},0)=0$, using \textbf{Theorem 1}, \textbf{Theorem 2}, and some well-known theorems of real analysis, we have proved the following theorems to confirm the existence of solitary wave and double layer solution of the energy integral (\ref{energy int}) at $M=M_{c}$.\\
\textbf{Theorem 3 :} If  $V'''(M_{c},0)<0$, then at $M=M_{c}$, there may exist either a solitary wave or a double layer in the positive potential side but there does not exist any solitary wave or double layer in the negative potential side.\\
\textbf{Theorem 4 :} If $V'''(M_{c},0)>0$, then at $M=M_{c}$, there may exist either a solitary wave or a double layer in the negative potential side but there does not exist any solitary wave or double layer in the positive potential side.\\
\textbf{Theorem 5 :} If $V'''(M_{c},0)\neq 0$, then it is not possible to have coexistence of both positive and negative potential solitary structures (including double layers) at $M=M_{c}$.\\
\textbf{Theorem 6 :} If $V'''(M_{c},0)<0$, $\partial V/\partial M<0$ for all $M>0$ and for all $\phi>0$, and if there exists at least one value $M_{0}$ of $M$ such that the system supports Positive Potential Solitary Waves (PPSWs) for all  $M_{c}<M< M_{0}$, then there exist either a PPSW or a Positive Potential Double Layer (PPDL) at $M=M_{c}$.\\
\textbf{Theorem 7 :} If $V'''(M_{c},0)>0$, $\partial V/\partial M<0$ for all $M>0$ and for all $\phi<0$, and if there exists at least one value $M_{0}$ of $M$ such that the system supports Negative Potential Solitary Waves (NPSWs) for all  $M_{c}<M< M_{0}$, then there exist either a NPSW or a Negative Potential Double Layer (NPDL) at $M=M_{c}$.\\
\textbf{Theorem 8 :} If $V'''(M_{c},0)<0$, $\partial V/\partial M<0$ for all $M>0$ and for all $\phi>0$, and if the system supports only NPSWs for $M>M_{c}$, then there does not exist any PPSW at $M=M_{c}$.\\
\textbf{Theorem 9 :} If $V'''(M_{c},0)>0$, $\partial V/\partial M<0$ for all $M>0$ and for all $\phi<0$, and if the system supports only PPSWs for $M>M_{c}$, then there does not exist any NPSW at $M=M_{c}$.\\
\textbf{Theorem 10 :} If $V'''(M_{c},0)\neq 0$, $\partial V/\partial M<0$ for all $M>0$ and for all $\phi \neq 0$, and if there exists at least one value $M_{0}$ of $M$ such that the system supports both PPSWs and NPSWs for all  $M_{c}<M< M_{0}$, i.e., if the system supports coexistence of both PPSWs and NPSWs for all $M_{c}<M< M_{0}$, then there exist either a PPSW or a PPDL at $M=M_{c}$ if $V'''(M_{c},0)<0$, whereas for $V'''(M_{c},0)>0$, there exist either a NPSW or a NPDL at $M=M_{c}$.

Apart from the conditions of \textbf{Theorem-6} (\textbf{Theorem-7}), the PPDL (NPDL) solution at $M=M_{c}$ is possible only when there exists a PPDL (NPDL) solution in any right neighborhood of $M_{c}$, i.e., PPDL (NPDL) solution at $M=M_{c}$ is possible only when the curve $M=M_{D}$ tends to intersect the curve $M=M_{c}$ at some point in the solution space of the energy integral, where each point of the curve $M=M_{D}$ corresponds to a PPDL (NPDL) solution of the energy integral whenever $M_{D}>M_{c}$. So, without going through all qualitatively different solution spaces of the energy integral by considering the entire range of parameters involved in the system, it is not possible to make a systematic investigation of the solitary wave solution and/or double layer solution of the energy integral at $M=M_{c}$. The most interesting fact is the existence of the double layer solution at $M=M_{c}$.

The existence of solitary structures at $M=M_{c}$ indicates that where and when one can apply small amplitude theory to find the weakly nonlinear solitary structures because of the following reasons:

(a) If in a portion of the solution space, there exist solitary waves or double layers of finite amplitude at $M=M_{c}$, then the amplitude of solitary waves corresponding to all $M>M_{c}$ is greater than that finite value. Then it is unwise to study weakly nonlinear solitary structures by means of any small amplitude methods (viz., replacing $V(\phi)$ by $V_{n}(\phi)$, where $V_{n}(\phi)$ represents the power series expansion of the Sagdeev potential or pseudo-potential $V(\phi)$ in $\phi$ keeping terms up to $\phi^{n}$) in that portion of the solution space, since these methods may lead to erroneous results. Actually, it is not possible to apply small amplitude method to study the weakly nonlinear solitary structures in a parameter regime where finite amplitude solitary structures exist at $M=M_{c}$, because in any small amplitude method, amplitude of solitary structure always tends to zero as $M \rightarrow M_{c}+0 \Leftrightarrow M > M_{c} \mbox{  and  } M \rightarrow M_{c}$ which contradicts the occurrence of finite amplitude solitary structure at $M=M_{c}$. Therefore, to study the weakly nonlinear solitary structures by means of small amplitude methods for $M>M_{c}$, the simple prescription is to investigate the nonlinear solitary structures at $M=M_{c}$. So, one cannot apply small amplitude theory to find weakly nonlinear solitary structures without investigating the existence of solitary structures at $M=M_{c}$ in the entire compositional parameter space.

(b) If in a portion of the solution space, there exist solitary structures having a finite amplitude at $M=M_{c}$, then it is simple to find the amplitude as well as width of the solitary structure for $M=M_{c}+M_{\epsilon}$ (for sufficiently small strictly positive quantity $M_{\epsilon}$) and this can be easily done by plotting $V(\phi)$ against $\phi$ at $M=M_{c}+M_{\epsilon}$. However, if in a portion of the solution space, there does not exist solitary structures at $M=M_{c}$, then it is very difficult to find the amplitude as well as width of the solitary wave for $M=M_{c}+M_{\epsilon}$ by plotting $V(\phi)$ against $\phi$. In this situation, the general prescription is to use the small amplitude method for finding the weakly nonlinear solitary structure at $M=M_{c}+M_{\epsilon}$.

(c) Considering any specific problem, it can be easily checked that the solitary structure along the curve $M=M_{c}$ is same as the solitary structure for $M>M_{c}$, if one can move the solution space through the family of curves parallel to the curve $M=M_{c}$.

Finally, we have used the analytical results as presented in this paper to a specific problem of dust acoustic wave in nonthermal plasma to investigate new results and new physical ideas.

The present paper is organized as follows: Physical interpretation of the energy integral for the existence of solitary wave and double layer solutions is given in Sec. \ref{sec:physics}. In Sec. \ref{sec:math}, some mathematical preliminaries (including definitions, properties and some basic theorems) have been given. Using these definitions, properties and basic theorems we have proved ten important theorems in Sec. \ref{sec:proof1-10} for the existence of solitary wave and/or double layer solution at $M=M_{c}$. The analytical results have been verified numerically by considering a problem of dust acoustic wave in nonthermal plasma in Sec. \ref{sec:verify}. Finally, a brief conclusion has been given in Sec. \ref{sec:conclusion}.

\section{\label{sec:physics}Physical interpretation of the energy integral}
The energy integral (\ref{energy int}) can be regarded as the one-dimensional motion of a particle of unit mass whose position is $\phi$ at time $\xi$ with velocity $d\phi/d\xi$ in a potential well $V(\phi)$. The first term of the energy integral can be regarded as the kinetic energy of a particle of unit mass at position $\phi$ and time $\xi$ whereas $V(\phi)$ is the potential energy of the same particle at that instant. Since kinetic energy is always a non-negative quantity, $V(\phi) \leq 0$ for the entire motion, i.e., zero is the maximum value for $V(\phi)$. Again from (\ref{energy int}), we find $d^{2} \phi/d \xi^{2} + V'(\phi) = 0$, i.e. the force acting on the particle at the position $\phi$ is $-V'(\phi)$. Suppose, $V(0) = V'(0) = 0$, therefore, the particle is in equilibrium at $\phi=0$ because the velocity as well as the force acting on the particle at $\phi=0$ are simultaneously equal to zero. Now if $\phi = 0$ can be made an unstable position of equilibrium, the energy integral can be interpreted as the motion of an oscillatory particle if $V(\phi_{m}) = 0$ for some $\phi_{m} \neq 0$, i.e., if the particle is slightly displaced from its unstable position of equilibrium then it moves away from its unstable position of equilibrium and it continues its motion until its velocity is equal to zero, i.e., until $\phi$ takes the value $\phi_{m}$. Now the force acting on the particle of unit mass at position $\phi = \phi_{m}$ is $-V'(\phi_{m})$. For $\phi_{m}<0$, the force acting on the particle at the point $\phi=\phi_{m}$ is directed towards the point $\phi=0$ if $-V'(\phi_{m})>0$, i.e., if $V'(\phi_{m})<0$. On the other hand, for $\phi_{m}>0$, the force acting on the particle at the point $\phi=\phi_{m}$ is directed towards the point $\phi=0$ if $-V'(\phi_{m})<0$, i.e., if $V'(\phi_{m})>0$. Therefore, if $V'(\phi_{m})>0$ (for the positive potential side ) or if $V'(\phi_{m})<0$ (for the negative potential side ) then the particle reflects back again to $\phi = 0$. Again, if $V(\phi_{m})=V'(\phi_{m})=0$ then the velocity $d \phi/d\xi$ as well as the force $d^{2} \phi/d \xi^{2}$ both are simultaneously equal to zero at $\phi = \phi_{m}$. Consequently, if the particle is slightly displaced from its unstable position of equilibrium ($\phi = 0$) it moves away from $\phi = 0$ and it continues its motion until the velocity is equal to zero, i.e., until $\phi$ takes the value $\phi = \phi_{m}$. However it cannot be reflected back again at $\phi = 0$ as the velocity and the force acting on the particle at $\phi = \phi_{m}$ vanish simultaneously. Actually, if $V'(\phi_{m})>0$ (for $\phi_{m}>0$) or if $V'(\phi_{m})<0$ (for $\phi_{m}<0$) the particle takes an infinite long time to move away from the unstable position of equilibrium. After that it continues its motion until $\phi$ takes the value $\phi_{m}$ and again it takes an infinite long time to come back its unstable position of equilibrium. Therefore, for the existence of a PPSW (NPSW) solution of the energy integral (\ref{energy int}), we must have the following: (a) $\phi=0$ is the position of unstable equilibrium of the particle, (b) $V(\phi_{m}) = 0$, $V'(\phi_{m}) > 0$ $(V'(\phi_{m}) < 0)$ for some $\phi_{m} > 0$ $(\phi_{m} < 0)$, which is nothing but the condition for oscillation of the particle within the interval $\min\{0,\phi_{m}\}<\phi<\max\{0,\phi_{m}\}$  and (c) $V(\phi) < 0$ for all $0 <\phi < \phi_{m}$ $(\phi_{m} < \phi < 0$), which is the condition to define the energy integral (\ref{energy int}) within the interval $\min\{0,\phi_{m}\}<\phi<\max\{0,\phi_{m}\}$. For the existence of a PPDL (NPDL) solution of the energy integral (\ref{energy int}), the conditions (a) and (c) remain unchanged but here (b) has been modified in such a way that the particle cannot be reflected again at $\phi = 0$, i.e., the condition (b) assumes the following form: $V(\phi_{m}) = V'(\phi_{m}) = 0$, $V''(\phi_{m}) < 0$ for some $\phi_{m} > 0$ $(\phi_{m} < 0$).

The above discussions for the existence of solitary waves and double layers are valid if $\phi = 0$ is an unstable position of equilibrium, i.e., if $V''(0) < 0$ along with $V(0) = V'(0) = 0$. In other words, $\phi=0$ can be made an unstable position of equilibrium if the potential energy of the said particle attains its maximum value at $\phi=0$. Now, the condition $V''(0)<0$ gives a lower bound $M_{c}$ of $M$, i.e., $V''(0)<0\Leftrightarrow M>M_{c}$, $V''(0)>0\Leftrightarrow M<M_{c}$, and $V''(0)=0\Leftrightarrow M=M_{c}\Leftrightarrow V''(M_{c},0)=0$. This $M_{c}$ is, in general, a function of the parameters involved in the system, or a constant. Therefore, if $M<M_{c}$, the potential energy of the said particle attains its minimum value at $\phi=0$, and consequently, $\phi=0$ is the position of stable equilibrium of the particle, and in this case, it is impossible to make any oscillation of the particle even when it is slightly displaced from its position of stable equilibrium, and consequently there is no question of existence of solitary waves or double layers for $M<M_{c}$. In other words, for the position of unstable equilibrium of the particle at $\phi=0$, i.e., for $V''(0)<0(\Leftrightarrow M>M_{c})$, the function $V(\phi)$ must be convex within a neighborhood of $\phi=0$ and in this case both type of solitary waves (negative or positive potential) may exist if other conditions are fulfilled. Now suppose that $V''(M_{c},0)=0$ and also $V'''(M_{c},0)=0$, then if $V '''' (M_{c},0)<0$, the potential energy of the said particle attains its maximum value at $\phi=0$ and consequently, $\phi=0$ is the position of unstable equilibrium. On the other hand if $V''(M_{c},0)=0$, $V'''(M_{c},0)=0$, and $V '''' (M_{c},0)>0$, the potential energy of the said particle attains its minimum value at $\phi=0$ and consequently, $\phi=0$ is the position of stable equilibrium of the particle and in this case there is no question of existence of solitary wave solution and/or double layer solution of the energy integral (\ref{energy int}). But if $V'''(M_{c},0)\neq 0$ along with $V(M_{c},0)=V'(M_{c},0)=V''(M_{c},0)=0$, then without going through the complete analytical investigation, it is difficult to predict the existence of solitary wave and/or double layer solution of the energy integral (\ref{energy int}) at $M=M_{c}$. In this situation, i.e., when $V'''(M_{c},0)\neq 0$ along with $V(M_{c},0)=V'(M_{c},0)=V''(M_{c},0)=0$, to confirm  the existence of solitary wave and/or double layer solution of the energy integral (\ref{energy int}) at $M=M_{c}$, some definitions and well-known theorems of real analysis are necessary. So, in the next section, we consider only those definitions and theorems of real analysis for continuation of physical interpretation of the existence of solitary wave and/or double layer solution of the energy integral (\ref{energy int}) at $M=M_{c}$ when $V'''(M_{c},0)\neq 0$ along with $V(M_{c},0)=V'(M_{c},0)=V''(M_{c},0)=0$.
\section{\label{sec:math}Mathematical preliminaries}
\subsection{\label{subsec:def}Definitions and properties} 
\textbf{D1 :} Let $a$ be any point of $\Re$, the set of all real numbers and $\delta >0$ is a strictly positive real number, then the $\delta-$ neighborhood of the point $a$ is denoted by $N(a,\delta)$ and is defined as a set of all real numbers $x$ such that $a-\delta<x< a+\delta$, i.e., $N(a,\delta)=(a-\delta, a+\delta)=\{x:a-\delta<x< a+\delta\}$, i.e.,
\begin{eqnarray}\label{open nbd}
x \in N(a,\delta) \Leftrightarrow a-\delta<x< a+\delta \Leftrightarrow |x-a|<\delta.
\end{eqnarray}
\textbf{P1 :} If $0<\delta_{1} \leq \delta_{2}$, using Eq. (\ref{open nbd}) it is simple to check the following property:
\begin{eqnarray}\label{nbd p1}
N(a,\delta_{1}) \subseteq N(a,\delta_{2}).
\end{eqnarray}
\textbf{D2 :} Let $A $ be any subset of $\Re$ and $a \in A$, then $a$ is said to be an interior point of $A$ if there exists a $\delta >0$ such that $N(a,\delta) \subseteq A$, i.e., any point of $N(a,\delta)$ is also a point of $A$.\\
\textbf{D3 :} $\Re \times \Re$ is the set of all two-dimensional points $(x,y)$, i.e., $\Re \times \Re$ is the set of all ordered pairs $(x,y)$ such that $x \in \Re$ and $y \in \Re$, i.e.,
\begin{eqnarray}\label{2D}
\Re \times \Re=\{(x,y):x \in \Re \mbox{  and  } y \in  \Re\}.
\end{eqnarray}
If $A$ and $B$ are any two subsets of $\Re$, then we define $A \times B$ as the set of all ordered pairs $(x,y)$ such that $x \in A$ and $y \in B$, i.e.,
\begin{eqnarray}\label{subset 2D}
A \times B=\{(x,y):x \in A \subseteq \Re \mbox{  and  } y \in  B \subseteq \Re\}.
\end{eqnarray}
\textbf{P2 :} If $A$, $B$, $C$ and $D$ are any four subsets of $\Re$ and $A \times B \subseteq C \times D$, then it is simple to prove the following property:
\begin{eqnarray}\label{subset 2D property}
A \subseteq C \mbox{  and  } B \subseteq D.
\end{eqnarray}
\textbf{P3 :} For any four subsets $A$, $B$, $C$ and $D$ of $\Re$, it is simple to prove the following property: $(A \times B) \cap (C \times D)=(A  \cap C) \times (B \cap D)$.\\
\textbf{D4 :} Let $(a,b)$ be any point of $\Re \times \Re$ and $\delta_{1} >0$, $\delta_{2} >0$ are two strictly positive real numbers, then the $\delta_{1}\delta_{2}$- rectangular neighborhood of the point $(a,b)$ is denoted by $N(a,b;\delta_{1},\delta_{2})$ and is defined as a set of all points $(x,y)$ of $\Re \times \Re$ such that $a-\delta_{1}<x< a+\delta_{1}$ and $b-\delta_{2}<y< b+\delta_{2}$, i.e., $ N(a,b;\delta_{1},\delta_{2})
=\{(x,y):x \in N(a,\delta_{1})\mbox{  and  }y \in N(b,\delta_{2})\}
=\{(x,y):a-\delta_{1}<x< a+\delta_{1} \mbox{  and  } b-\delta_{2}<y< b+\delta_{2}\}
=\{(x,y):|x-a|< \delta_{1} \mbox{  and  } |y-b|< \delta_{2}\}$, i.e.,
\begin{eqnarray}\label{open nbd 2D}
& &(x,y) \in N(a,b;\delta_{1},\delta_{2}) \nonumber \\
&\Leftrightarrow& |x-a|< \delta_{1} \mbox{  and  } |y-b|< \delta_{2}.
\end{eqnarray}
\textbf{P4 :} $N(a,b;\delta_{1},\delta_{2})=N(a,\delta_{1})\times N(b,\delta_{2})$\\
\textbf{D5 :} Let $A \times B$ be any subset of $\Re \times \Re$ and $(a,b)$ be a point of $A \times B$, i.e., $a \in A$ and $b \in B$, then $(a,b)$ is said to be an interior point of $A \times B$ if there exists a $\delta_{1}\delta_{2}$- rectangular neighborhood of the point $(a,b)$ with $\delta_{1} >0$, $\delta_{2} >0$ such that $N(a,b;\delta_{1},\delta_{2}) \subseteq A \times B$, i.e., any point of $N(a,b;\delta_{1},\delta_{2})$ is also a point of $A \times B$, i.e., $N(a,\delta_{1}) \times N(b,\delta_{2}) \subseteq A \times B$, i.e., $N(a,\delta_{1}) \subseteq A$ and $N(b,\delta_{2}) \subseteq  B$.
\subsection{\label{bth}Some theorems of real analysis}
\textbf{BTh1 :} Let $f(x)$ be a real valued function defined on a subset $D$ of the set of real numbers $\Re$ and $f(x)$ be continuous at $x=a \in D$. If $f(a) \neq 0$, then there exists a real number $\delta_{1}>0$ such that $f(a)f(x)>0$ for all $x \in N(a,\delta_{1})\cap D$. Furthermore, if $a$ is an interior point of D, then there exists a real number $\delta>0$ such that $f(a)f(x)>0$ for all $x \in N(a, \delta)\subseteq D$.\\
\textbf{BTh2 :} Let $f(x,y)$ be a real valued function defined on a subset $D=A\times B$ of $\Re\times\Re$ and $f(x,y)$ be continuous at $(x,y)=(a,b) \in D$. If $f(a,b) \neq 0$, then there exist two real numbers $\delta_{1}>0$ and $\delta_{2}>0$ such that $f(a,b)f(x,y)>0$ for all $(x,y) \in N(a,b;\delta_{1},\delta_{2})\cap D$. Furthermore, if $(a,b)$ is an interior point of $D$, then there exist two real numbers $\delta_{3}>0$ and $\delta_{4}>0$ such that $f(a,b)f(x,y)>0$ for all $(x,y) \in N(a,b;\delta_{3},\delta_{4})\subseteq D$.\\
\textbf{Bolzano's Theorem :} Let $f(x)$ be a real valued function defined on a closed interval $[a,b]$ ($a < b$). If $f(x)$ is continuous on $[a,b]$ and $f(a)f(b)<0$, then there exists a point $c \in (a,b)$ such that $f(c)=0$.\\
\textbf{Lagrange's Mean - Value Theorem :} Let $f(x)$ be a real valued function defined on a closed interval $[a,b]$ ($a < b$). If $f(x)$ is continuous on $[a,b]$ and if it is derivable on $(a,b)$, then there exists a point $c \in (a,b)$ such that
\begin{eqnarray}
	[f(b)-f(a)]/(b-a)=f'(c),\nonumber
\end{eqnarray}
where `` $'$ '' indicates derivative with respect to $x$.\\
\textbf{IFT (The implicit function theorem of two variables) :} Let $V(M,\phi)$ be a real valued function of two variables $M$ and $\phi$ defined on $\mathcal{M}$ $\times$ $\Phi$, i.e., for each $M \in \mathcal{M}$ and for each $\phi \in \Phi $, there exists a unique real number $V(M,\phi)$. Let $(M_{1},\phi_{1}) \in \mathcal{M} \times \Phi$ be an interior point of $\mathcal{M} \times \Phi$ and\\
\textbf{A1 :} $V(M_{1},\phi_{1})=0$,\\
\textbf{A2 :} $\frac{\partial V}{\partial \phi}$ is continuous in a rectangular neighborhood of $(M_{1},\phi_{1})$,\\
\textbf{A3 :} $\frac{\partial V}{\partial \phi}\bigg|_{(M_{1},\phi_{1})} \neq 0$,\\
there exists a rectangular neighborhood $N(M_{1},\phi_{1};h_{1},k_{1})=N(M_{1},h_{1}) \times N(\phi_{1},k_{1})$ of $(M_{1},\phi_{1})$ such that corresponding to every $M \in N(M_{1},h_{1})$, $\phi (\in N(\phi_{1},k_{1}))$ can be expressed uniquely as a function of $M$, for which the following conditions are simultaneously satisfied.\\
\textbf{C1 :} $\phi(M_{1})=\phi_{1}$,\\
\textbf{C2 :} $V(M,\phi(M))=0$ for all $M \in N(M_{1},h_{1})$,\\
\textbf{C3 :} $\frac{d \phi}{d M}=-(\frac{\partial V}{\partial M} \div \frac{\partial V}{\partial \phi})$ for all $(M, \phi) \in N(M_{1},\phi_{1};h_{1},k_{1})=N(M_{1},h_{1}) \times N(\phi_{1},k_{1})$.

To prove Theorem 1 - Theorem 10, it is helpful to remember that if a property is true for each point of a set then that property is also true for each point of any subset of that set. Without any loss of generality, we can assume $(M, \phi)$ as an interior point of $\mathcal{M} \times \Phi$.
\section{\label{sec:proof1-10}Proof of Theorem 1 - Theorem 10}
\textbf{Theorem 1 :} If $V'''(M_{c},0)\neq 0$, then there exists a strictly positive real number $\phi_{\epsilon}$ ($>0$) such that $V'''(M_{c},0) V'''(M_{c},\phi) > 0$ for all $-\phi_{\epsilon} < \phi < \phi_{\epsilon}$.\\
\textbf{Proof :} Here $f(\phi)=V'''(M_{c},\phi)$ is continuous at $\phi=0$ and consequently from \textbf{BTh1}, we can conclude that there exists a strictly positive real number $\phi_{\epsilon}$ ($>0$) such that
\begin{eqnarray}
& &	f(\phi)f(0)>0 ~~\forall~~ \phi \in (0-\phi_{\epsilon}, 0+\phi_{\epsilon}),\nonumber \\
&\Rightarrow& V'''(M_{c},\phi)V'''(M_{c},0)>0 ~~\forall~~ \phi \in (-\phi_{\epsilon}, \phi_{\epsilon}),\nonumber \\
&\Rightarrow& V'''(M_{c},\phi)V'''(M_{c},0)>0 ~~\forall~~  -\phi_{\epsilon}<\phi< \phi_{\epsilon}.\nonumber \\
\end{eqnarray}
\textbf{Theorem 2 :} If $V''(M_{c},0) = 0$ and $V'''(M_{c},0)\neq 0$, then there exists a strictly positive real number $\phi_{\epsilon}$ ($>0$) such that $V'''(M_{c},0) V''(M_{c},\phi) > 0$ for all $0 < \phi < \phi_{\epsilon}$ and $V'''(M_{c},0) V''(M_{c},\phi) < 0$ for all $-\phi_{\epsilon} < \phi < 0$.\\
\textbf{Proof :} From \textbf{Theorem 1}, we see that if $V'''(M_{c},0)\neq 0$, then there exists a strictly positive real number $\phi_{\epsilon}$ ($>0$) such that
\begin{eqnarray}
	V'''(M_{c},\phi)V'''(M_{c},0)>0 ~~\forall~~  -\phi_{\epsilon}<\phi< \phi_{\epsilon}.\label{theorem1}
\end{eqnarray}
Again as $V(M_{c},\phi)$ and its derivatives of all order with respect to $\phi$ exist finitely, $f(\phi)=V''(M_{c},\phi)$ is a continuously derivable function of $\phi \in \Phi$. Now, for any $\phi (\neq 0) \in (-\phi_{\epsilon},\phi_{\epsilon})$, we have (a) the real valued function $f(\phi)$ is well-defined in the closed interval [$\min\{0,\phi\}$, $\max\{0,\phi\}$], (b) the real valued function $f(\phi)$ is continuous in the closed interval [$\min\{0,\phi\}$, $\max\{0,\phi\}$], (c) the real valued function $f(\phi)$ is derivable in the open interval ($\min\{0,\phi\}$, $\max\{0,\phi\}$). Therefore, from the Lagrange's mean - value theorem of differential calculus, we can conclude that there exists a $\phi_{1} \in (\min\{0,\phi\},\max\{0,\phi\})$ such that
\begin{eqnarray}
&& \frac{f(\max\{0,\phi\})-f(\min\{0,\phi\})}{\max\{0,\phi\}-\min\{0,\phi\}}=f'(\phi_{1}),\nonumber \\
&&\Rightarrow \frac{f(\phi)-f(0)}{\phi-0}=\frac{f(0)-f(\phi)}{0-\phi}=f'(\phi_{1}).\label{lagrange}
\end{eqnarray}
As $f(0)=V''(M_{c},0) = 0$ and $\phi \neq 0$, (\ref{lagrange}) assumes the following form:
\begin{eqnarray}
&& \frac{V''(M_{c},\phi)}{\phi} = V'''(M_{c},\phi_{1}),\nonumber \\
\Rightarrow && V''(M_{c},\phi) = \phi V'''(M_{c},\phi_{1}),\nonumber \\
\Rightarrow && V'''(M_{c},0)V''(M_{c},\phi)
 = \phi V'''(M_{c},0)V'''(M_{c},\phi_{1}).\label{v2dash}
\end{eqnarray}
Suppose that $\phi<0$, then $\min\{0,\phi\}=\phi$ and $\max\{0,\phi\}=0$, and consequently, $\min\{0,\phi\}<\phi_{1}<\max\{0,\phi\} \Rightarrow \phi<\phi_{1}<0$, but since $\phi (\neq 0) \in (-\phi_{\epsilon},\phi_{\epsilon})$ we have $-\phi_{\epsilon}<\phi<\phi_{\epsilon}\Rightarrow -\phi_{\epsilon}<\phi<\phi_{1}<0<\phi_{\epsilon}\Rightarrow -\phi_{\epsilon}<\phi_{1}<\phi_{\epsilon} \Rightarrow \phi_{1} \in (-\phi_{\epsilon},\phi_{\epsilon})$.

Next suppose that $\phi>0$, then $\min\{0,\phi\}=0$ and $\max\{0,\phi\}=\phi$, and consequently, $\min\{0,\phi\}<\phi_{1}<\max\{0,\phi\} \Rightarrow 0<\phi_{1}<\phi$, but since $\phi (\neq 0) \in (-\phi_{\epsilon},\phi_{\epsilon})$ we have $-\phi_{\epsilon}<\phi<\phi_{\epsilon}\Rightarrow -\phi_{\epsilon}<0<\phi_{1}<\phi<\phi_{\epsilon}\Rightarrow -\phi_{\epsilon}<\phi_{1}<\phi_{\epsilon} \Rightarrow \phi_{1} \in (-\phi_{\epsilon},\phi_{\epsilon})$.

Therefore in any case ($\phi<0$ or $\phi>0$), we always have $\phi_{1} \in (-\phi_{\epsilon},\phi_{\epsilon})$, and consequently, from Eq. (\ref{theorem1}), we get
\begin{eqnarray}
	V'''(M_{c},\phi_{1})V'''(M_{c},0)>0 .\label{v3dash10}
\end{eqnarray}
From Eq. (\ref{v3dash10}), we get
\begin{eqnarray}
&&\phi V'''(M_{c},\phi_{1})V'''(M_{c},0)<0 \forall -\phi_{\epsilon}<\phi<0, \label{phiv3dash10 lt 0}\\
&&\phi V'''(M_{c},\phi_{1})V'''(M_{c},0)>0 \forall~ 0<\phi<\phi_{\epsilon}.\label{phiv3dash gt 0}
\end{eqnarray}
From Eq. (\ref{v2dash}), using Eqs. (\ref{phiv3dash10 lt 0}) and (\ref{phiv3dash gt 0}), we get
\begin{eqnarray}
&& V'''(M_{c},0)V''(M_{c},\phi)<0 ~~\forall~~ -\phi_{\epsilon}<\phi<0 ,\\
&& V'''(M_{c},0)V''(M_{c},\phi)>0 ~~\forall~~ 0<\phi<\phi_{\epsilon},
\end{eqnarray}
and consequently, Theorem 2 is proved.

From Theorem 2, we observe that whenever $V(M,0)=V'(M,0)=V''(M_{c},0)=0$ and $V'''(M_{c},0)<0$, then $V''(M_{c},\phi)>0$ for all $-\phi_{\epsilon} < \phi < 0$ and $V''(M_{c},\phi)<0$ for all $0 < \phi < \phi_{\epsilon}$, i.e., $V(M_{c},\phi)$ is a concave function within the interval $-\phi_{\epsilon} < \phi < 0$ and $V(M_{c},\phi)$ is a convex function within the interval $0 < \phi < \phi_{\epsilon}$, where $\phi_{\epsilon}$ is strictly positive quantity but it can be made sufficiently small. Therefore, if the particle placed at $\phi=0$ be slightly displaced towards the positive potential side, it falls within the interval $0 < \phi < \phi_{\epsilon}$ and due to the convexity of the function $V(M_{c},\phi)$ within the interval $0 < \phi < \phi_{\epsilon}$, it moves away from $\phi = 0$ and it continues its motion until its velocity is equal to zero, i.e., until $\phi$ takes the value $\phi = \phi_{c}>0$, where $V(M_{c},\phi_{c})=0$ and in this case if $V'(M_{c},\phi_{c}) > 0$ ($V'(M_{c},\phi_{c}) = 0$ and $V''(M_{c},\phi_{c}) < 0$), one can easily get a PPSW (PPDL) as a solution of the energy integral (\ref{energy int}). Again, as $V(M_{c},\phi)$ is a concave function within the interval $-\phi_{\epsilon} < \phi < 0$, if the particle placed at $\phi=0$ be slightly displaced towards the negative potential side, it falls within the interval $-\phi_{\epsilon} < \phi < 0$ and due to the concavity of the function $V(M_{c},\phi)$ within the interval $-\phi_{\epsilon} < \phi < 0$, it moves towards the point $\phi = 0$ and consequently, there does not exist any solitary wave or double layer solution in the negative potential side. In true physical sense, $\lim_{\phi \rightarrow 0+0}V(M_{c},\phi)$ is an unstable position which lies in a small right neighborhood of the equilibrium position $\phi=0$, whereas $\lim_{\phi \rightarrow 0-0}V(M_{c},\phi)$ is a stable position which lies in a small left neighborhood of the equilibrium position $\phi=0$, where $\phi \rightarrow 0+0 \Leftrightarrow \phi >0$ and $\phi$ approaches to zero, i.e., $\phi$ approaches to zero from the right side of zero and  $\phi \rightarrow 0-0 \Leftrightarrow \phi <0$ and $\phi$ approaches to zero, i.e., $\phi$ approaches to zero from the left side of zero. Hence, as $\lim_{\phi \rightarrow 0+0}V(M_{c},\phi)$ is an unstable position which lies in a small right neighborhood of the equilibrium position $\phi=0$, there may exist either a solitary wave or a double layer in the positive potential side. On the other hand, as $\lim_{\phi \rightarrow 0-0}V(M_{c},\phi)$ is a stable position which lies in a small left neighborhood of the equilibrium position $\phi=0$, there does not exist any solitary wave or double layer in the negative potential side.  Hence the above theoretical discussions can be summarized as follows.\\
\textbf{Theorem 3 :} If $V(M,0)=V'(M,0)=V''(M_{c},0)=0$ and $V'''(M_{c},0)<0$, then at $M=M_{c}$, there may exist either a solitary wave or a double layer in the positive potential side but there does not exist any solitary wave or double layer in the negative potential side.

Again from Theorem 2, we observe that whenever $V(M,0)=V'(M,0)=V''(M_{c},0)=0$ and $V'''(M_{c},0)>0$, then $V''(M_{c},\phi)<0$ for all $-\phi_{\epsilon} < \phi < 0$ and $V''(M_{c},\phi)>0$ for all $0 < \phi < \phi_{\epsilon}$, i.e., $V(M_{c},\phi)$ is a convex function within the interval $-\phi_{\epsilon} < \phi < 0$ and $V(M_{c},\phi)$ is a concave function within the interval $0 < \phi < \phi_{\epsilon}$. Therefore, following the same argument as given in the proof of Theorem 3, we have\\
\textbf{Theorem 4 :} If $V(M,0)=V'(M,0)=V''(M_{c},0)=0$ and $V'''(M_{c},0)>0$, then at $M=M_{c}$, there may exist either a solitary wave or a double layer in the negative potential side but there does not exist any solitary wave or double layer in the positive potential side.

From Theorem 3 and Theorem 4, we note that if $V(0)=V'(0)=0, V''(M_{c},0)=0$ and $V'''(M_{c},0) \neq 0$, the point $\phi=0$ is the point of inflexion that seperates the convex part of $V(M_{c},\phi)$ from the concave part and if $\phi = 0$ is a point of inflexion then there may exist either a solitary wave or a double layer in either positive potential side or negative potential side, but the solitary structures (including double layers) of both polarities can not coexist at $M=M_{c}$. Therefore,\\
\textbf{Theorem 5 :} If $V(M,0)=V'(M,0)=V''(M_{c},0)=0$ and $V'''(M_{c},0)\neq 0$, it is not possible to have coexistence of both positive and negative potential solitary structures (including double layers) at $M=M_{c}$.

Next to confirm the existence of the solitary wave and/or double layer solution of the energy integral (\ref{energy int}) at $M=M_{c}$, we have proved the following theorems.\\
\textbf{Theorem 6 :} If $V(M,0)=V'(M,0)=V''(M_{c},0)=0$, $V'''(M_{c},0)<0$, $\partial V/\partial M<0$ for all $M>0$ and for all $\phi>0$, and if there exists at least one value $M_{0}$ of $M$ such that the system supports Positive Potential Solitary Waves (PPSWs) for all  $M_{c}<M< M_{0}$, then there exist either a PPSW or a Positive Potential Double Layer (PPDL) at $M=M_{c}$.\\
\textbf{Proof :} From \textbf{Theorem 3}, we have seen that if $V(M,0)=V'(M,0)=V''(M_{c},0)=0$ and $V'''(M_{c},0)<0$, there may exist either a solitary wave or a double layer in the positive potential side at $M=M_{c}$ but there does not exist any solitary wave or double layer in the negative potential side at $M=M_{c}$. Now with the help of IFT, we shall confirm the existence of either a solitary wave or a double layer in the positive potential side at $M=M_{c}$. Let $M_{c}<M_{1}< M_{0}$, then according to the condition of the present theorem, there exists a solitary wave in the positive potential side at $M=M_{1}$. Let the amplitude of this solitary wave is $\phi_{1} (>0)$, then from the conditions for existence of PPSW we have
\begin{description}
	\item[SM1-1 :] $V(M_{1},\phi_{1})=0$,
	\item[SM1-2 :] $\frac{\partial V}{\partial \phi}\bigg|_{(M_{1},\phi_{1})}>0$,
	\item[SM1-3 :] $V(M_{1},\phi)<0$ for all $0<\phi<\phi_{1}$.
\end{description}
Again, as $\partial V/\partial \phi$ is a continuous function of $(M,\phi)$, we get
\begin{description}
	\item[SM1-4 :] $\frac{\partial V}{\partial \phi}$ is continuous on a rectangular neighborhood of $(M_{1},\phi_{1})$.
\end{description}
Therefore, \textbf{SM1-1}, \textbf{SM1-4}, and \textbf{SM1-2} are, respectively, same as the conditions \textbf{A1}, \textbf{A2}, and \textbf{A3} of IFT, and consequently, we can apply IFT in a rectangular neighborhood of the point $(M_{1},\phi_{1})$. Therefore, from IFT, we can conclude that there exists a rectangular neighborhood $N(M_{1},\phi_{1};h_{1},k_{1})=N(M_{1},h_{1}) \times N(\phi_{1},k_{1})$ of $(M_{1},\phi_{1})$ such that corresponding to every $M \in N(M_{1},h_{1})$, $\phi (\in N(\phi_{1},k_{1}))$ can be expressed uniquely as a function of $M$, for which the following conditions are simultaneously satisfied.
\begin{description}
	\item[IFT-C1 :] $\phi(M_{1})=\phi_{1}$,
	\item[IFT-C2 :] $V(M,\phi(M))=0$ for all $M \in N(M_{1},h_{1})$,
	\item[IFT-C3 :] $\frac{d \phi}{d M}=-(\frac{\partial V}{\partial M} \div \frac{\partial V}{\partial \phi})$ for all $(M,\phi) \in N(M_{1},\phi_{1};h_{1},k_{1})=N(M_{1},h_{1}) \times N(\phi_{1},k_{1})$.
\end{description}

Again from \textbf{SM1-2}, using theorem \textbf{BTh2}, we have strictly positive real numbers $h_{2}(>0)$ and $k_{2}(>0)$ such that
\begin{description}
	\item[SM1-5 :] $\frac{\partial V}{\partial \phi}>0$ for all $(M,\phi) \in N(M_{1},\phi_{1};h_{2},k_{2})=N(M_{1},h_{2}) \times N(\phi_{1},k_{2})$.
\end{description}
Again from \textbf{IFT-C3}, we see that $\phi$ is derivable function of $M$ for $M \in (M_{1}-h_{1},M_{1}+h_{1})$, and consequently, it is continuous in $(M_{1}-h_{1},M_{1}+h_{1})$. As $\phi$ is continuous in $(M_{1}-h_{1},M_{1}+h_{1})$ and $\phi(M_{1})=\phi_{1}>0$, using theorem \textbf{BTh1}, we have a strictly positive real number $h_{3}(>0)$ such that
\begin{description}
	\item[SM1-6 :] $\phi(M)>0$ for all $M \in N(M_{1},h_{3})$.
\end{description}

So, if we take $h=\min \{h_{1}, h_{2}, h_{3}\}$ and $k=\min \{k_{1}, k_{2}\}$, then $N(M_{1},h) \subseteq N(M_{1},h_{i})$ for all $i=1,2,3$, and $N(\phi_{1},k)\subseteq N(\phi_{1},k_{j})$ for all $j=1,2$, consequently, $N(M_{1},\phi_{1};h,k)=N(M_{1},h) \times N(\phi_{1},k) \subseteq N(M_{1},h_{i}) \times N(\phi_{1},k_{j}) = N(M_{1},\phi_{1};h_{i},k_{j})$ for all $i=1,2,3$ and $j=1,2$. Therefore, we can also apply IFT in the rectangular neighborhood  $N(M_{1},\phi_{1};h,k)=N(M_{1},h) \times N(\phi_{1},k)$, and consequently, with respect to the rectangular neighborhood $N(M_{1},\phi_{1};h,k)$, \textbf{IFT-C1}, \textbf{IFT-C2}, \textbf{IFT-C3}, \textbf{SM1-5}, and \textbf{SM1-6} can be put in the following form:
\begin{description}
  \item[mSM1-1 :] $\phi(M_{1})=\phi_{1}$,
	\item[mSM1-2 :] $V(M,\phi(M))=0$ for all $M \in N(M_{1},h)$,
	\item[mSM1-3 :] $\frac{d \phi}{d M}=-(\frac{\partial V}{\partial M} \div \frac{\partial V}{\partial \phi})$ for all $(M,\phi) \in N(M_{1},\phi_{1};h,k)=N(M_{1},h) \times N(\phi_{1},k)$,
	\item[mSM1-4 :] $\frac{\partial V}{\partial \phi}>0$ for all $(M,\phi) \in N(M_{1},\phi_{1};h,k)=N(M_{1},h) \times N(\phi_{1},k)$,
	\item[mSM1-5 :] $\phi(M)>0$ for all $M \in N(M_{1},h)$.
\end{description}

Now choose two strictly positive real numbers $\epsilon_{1}$ and $\epsilon$ such that $M_{1}=M_{c}+\epsilon_{1}$ and $N(M_{c},\epsilon) \subseteq N(M_{1},h)$. These two conditions are simultaneously satisfied if we choose $0<\epsilon_{1}<h$ and $0<\epsilon<h-\epsilon_{1}$. As $\phi(M) \in N(\phi_{1},k)$ for every $M  \in N(M_{1},h)$ and as $M_{c} \in  N(M_{1},h)$, then we must have $\phi(M_{c}) \in N(\phi_{1},k)$ $\Rightarrow$ $\phi_{c} \in N(\phi_{1},k)$, where we set $\phi_{c}=\phi(M_{c})$. Now choose two strictly positive real number $\delta_{1}$ and $\delta$ such that $\phi_{1}=\phi_{c}+\delta_{1}$ and $N(\phi_{c},\delta) \subseteq N(\phi_{1},k)$. These two conditions are simultaneously satisfied if we choose $0<\delta_{1}<k$ and $0<\delta<k-\delta_{1}$.

Therefore, $N(M_{c},\epsilon) \subseteq N(M_{1},h)$ and $N(\phi_{c},\delta) \subseteq N(\phi_{1},k)$ $\Rightarrow$ $N(M_{c},\epsilon) \times N(\phi_{c},\delta) \subseteq N(M_{1},h) \times N(\phi_{1},k)$ $\Rightarrow$ $N(M_{c},\phi_{c};\epsilon,\delta) \subseteq N(M_{1},\phi_{1};h,k)$, and consequently, we can also apply IFT in the rectangular neighborhood  $N(M_{c},\phi_{c};\epsilon,\delta)\subseteq N(M_{1},\phi_{1};h,k)$. So, with respect to the rectangular neighborhood $N(M_{c},\phi_{c};\epsilon,\delta)$, \textbf{IFT-C1}, \textbf{IFT-C2}, \textbf{IFT-C3}, \textbf{mSM1-4}, and \textbf{mSM1-5} can be put in the following form:
\begin{description}
  \item[fmSM1-1 :] $\phi(M_{c})=\phi_{c}$,
	\item[fmSM1-2 :] $V(M,\phi(M))=0$ for all $M \in N(M_{c},\epsilon)$,
	\item[fmSM1-3 :] $\frac{d \phi}{d M}=-(\frac{\partial V}{\partial M} \div \frac{\partial V}{\partial \phi})$ for all $(M,\phi) \in N(M_{c},\phi_{c};\epsilon,\delta)=N(M_{c},\epsilon) \times N(\phi_{c},\delta)$,
	\item[fmSM1-4 :] $\frac{\partial V}{\partial \phi}>0$ for all $(M,\phi) \in N(M_{c},\phi_{c};\epsilon,\delta)=N(M_{c},\epsilon) \times N(\phi_{c},\delta)$,
	\item[fmSM1-5 :] $\phi(M)>0$ for all $M \in N(M_{c},\epsilon)$.
\end{description}
As $M_{c} \in N(M_{c},\epsilon)$ and $\phi_{c} \in N(\phi_{c},\delta)$, from \textbf{fmSM1-1}, \textbf{fmSM1-2}, \textbf{fmSM1-4}, and \textbf{fmSM1-5}, we get the following conditions:
\begin{description}
  \item[SMc-1 :] $V(M_{c},\phi_{c})=0$, where $\phi_{c}=\phi(M_{c})$,
	\item[SMc-2 :] $\frac{\partial V}{\partial \phi}\bigg|_{(M_{c},\phi_{c})}>0$,
	\item[SMc-3 :] $\phi_{c}=\phi(M_{c})>0$.
\end{description}

It is important to note that if we take $M \in (M_{c}-\epsilon,M_{c})$, then also the above three conditions are simultaneously satisfied but $M < M_{c} \Leftrightarrow V''(M,0)>0$, and consequently, $\phi=0$ is a position of stable equilibrium. So, there is no question of existence of any solitary wave or double layer solution of the energy integral (\ref{energy int}) for all $M \in (M_{c}-\epsilon,M_{c})$.

Again, from the condition of the present theorem, we have $\partial V/\partial M<0$ for all $M>0$ and for all $\phi> 0$. As our discussion is restricted to positive potential side ($\phi >0$) only, from \textbf{fmSM1-3} and \textbf{fmSM1-4}, we get
\begin{description}
  \item[Mc :] $\frac{d \phi}{d M} >0$ for all $M \in N(M_{c},\epsilon)$,
\end{description}
which shows that $\phi$ is an strictly increasing function of $M$ for all $M_{c}-\epsilon<M<M_{c}+\epsilon$.

Now we shall prove that
\begin{description}
	\item[SMc-4 :] $V(M_{c},\phi)\leq 0$ for all $0<\phi<\phi_{c}$,
\end{description}
and the following analysis have been developed to prove \textbf{SMc-4}.

Consider a value $M_{2}$ of $M$ such that $M_{c}<M_{2}<M_{c}+\epsilon$, then there exists a PPSW of amplitude $\phi_{2}$, and consequently we have
\begin{description}
	\item[SM2-1 :] $V(M_{2},\phi_{2})=0$,
	\item[SM2-2 :] $\frac{\partial V}{\partial \phi} \bigg|_{(M_{2},\phi_{2})}>0$,
	\item[SM2-3 :] $V(M_{2},\phi)<0$ for all $0<\phi<\phi_{2}$.
\end{description}
As $M_{c}<M_{2}<M_{c}+\epsilon \Rightarrow M_{c}-\epsilon<M_{2}<M_{c}+\epsilon$, we have $\phi_{2}=\phi(M_{2})$. Again as $\phi$ is an increasing function of $M$ for all $M_{c}-\epsilon<M<M_{c}+\epsilon$, and $M_{c}, M_{2} \in N(M_{c},\epsilon)$ with $M_{c}<M_{2}$, we have $\phi(M_{c})<\phi(M_{2})\Rightarrow \phi_{c}<\phi_{2}$. we can write \textbf{SM2-3} as follows
\begin{description}
	\item[mSM2-3 :] $V(M_{2},\phi)<0$ for all $0<\phi<\phi_{c}<\phi_{2}$.
\end{description}

If possible, let there exists a value $\phi_{p}$ of $\phi$ such that $0<\phi_{p}<\phi_{c}$ and
\begin{description}
	\item[Q1 :] $V(M_{c},\phi_{p})>0$.
\end{description}
Again from \textbf{mSM2-3}, as $0<\phi_{p}<\phi_{c}<\phi_{2}$, we get
\begin{description}
	\item[Q2 :] $V(M_{2},\phi_{p})<0$.
\end{description}
Considering $V(M, \phi_{p})$ as a function of $M$ only, from \textbf{Q1}, \textbf{Q2} and \textbf{Bolzano's Theorem}, we can conclude that there exists a value $M_{3}$ of $M$ such that 
\begin{eqnarray}\label{M3p30}
V(M_{3}, \phi_{p})=0,
\end{eqnarray}
where $M_{c}<M_{3}<M_{2}<M_{c}+\epsilon$. But as $M_{c}<M_{3}<M_{2}<M_{c}+\epsilon$, there exists a PPSW at $M=M_{3}$ of amplitude $\phi_{3}>0$, and consequently we have
\begin{description}
	\item[SM3-1 :] $V(M_{3},\phi_{3})=0$,
	\item[SM3-2 :] $\frac{\partial V}{\partial \phi} \bigg|_{(M_{3},\phi_{3})}>0$,
	\item[SM3-3 :] $V(M_{3},\phi)<0$ for all $0<\phi<\phi_{3}$,
\end{description}
As $M_{c}<M_{3}<M_{c}+\epsilon \Rightarrow M_{c}-\epsilon<M_{3}<M_{c}+\epsilon$, we have $\phi_{3}=\phi(M_{3})$. Again as $\phi$ is an increasing function of $M$ for $M_{c}-\epsilon<M<M_{c}+\epsilon$, and $M_{c}, M_{3} \in N(M_{c},\epsilon)$ with $M_{c}<M_{3}$, we have $\phi(M_{c})<\phi(M_{3})\Rightarrow \phi_{c}<\phi_{3}$. So we can write \textbf{SM3-3} as follows
\begin{description}
	\item[mSM3-3 :] $V(M_{3},\phi)<0$ for all $0<\phi<\phi_{c}<\phi_{3}$,
\end{description}
Again from \textbf{mSM3-3}, as $0<\phi_{p}<\phi_{c}<\phi_{3}$, we get
\begin{description}
	\item[Q3 :] $V(M_{3},\phi_{p})<0$,
\end{description}
which contradicts Eq. (\ref{M3p30}), i.e., $V(M_{3}, \phi_{p})=0$. Therefore, our supposition is wrong, and consequently, there does not exist any $\phi_{p} \in (0,\phi_{c})$ such that $V(M_{c},\phi_{p})>0$, i.e., $V(M_{c},\phi_{p})\ngtr 0$ and consequently, we get
\begin{description}
  \item[mSMc-4 :] $V(M_{c},\phi)\leq 0$ for all $0 < \phi < \phi_{c}$.
\end{description}

Now \textbf{mSMc-4} suggests the following two possible cases:
\begin{description}
  \item[Case - 1] There does not exist any $\phi \in (0,\phi_{c})$ such that $V(M_{c},\phi)= 0$, i.e., $V(M_{c},\phi)< 0$ for all $0 < \phi < \phi_{c}$.
  \item[Case - 2] There exists at least one $\phi \in (0,\phi_{c})$ such that $V(M_{c},\phi)= 0$.
\end{description}

Combining \textbf{Case - 1} with \textbf{SMc-1}, \textbf{SMc-2} and \textbf{SMc-3}, we get
\begin{description}
  \item[(1)] $V(M_{c},\phi_{c})=0$, where $\phi_{c}=\phi(M_{c})$,
	\item[(2)] $\frac{\partial V}{\partial \phi} \bigg|_{(M_{c},\phi_{c})}>0$,
	\item[(3)] $\phi_{c}=\phi(M_{c})>0$,
	\item[(4)] $V(M_{c},\phi)<0$ for all $0<\phi<\phi_{c}$.
\end{description}
These four conditions confirm the existence of PPSW solution of the energy integral (\ref{energy int}) at $M=M_{c}$.

\begin{figure}
\begin{center}
  \includegraphics{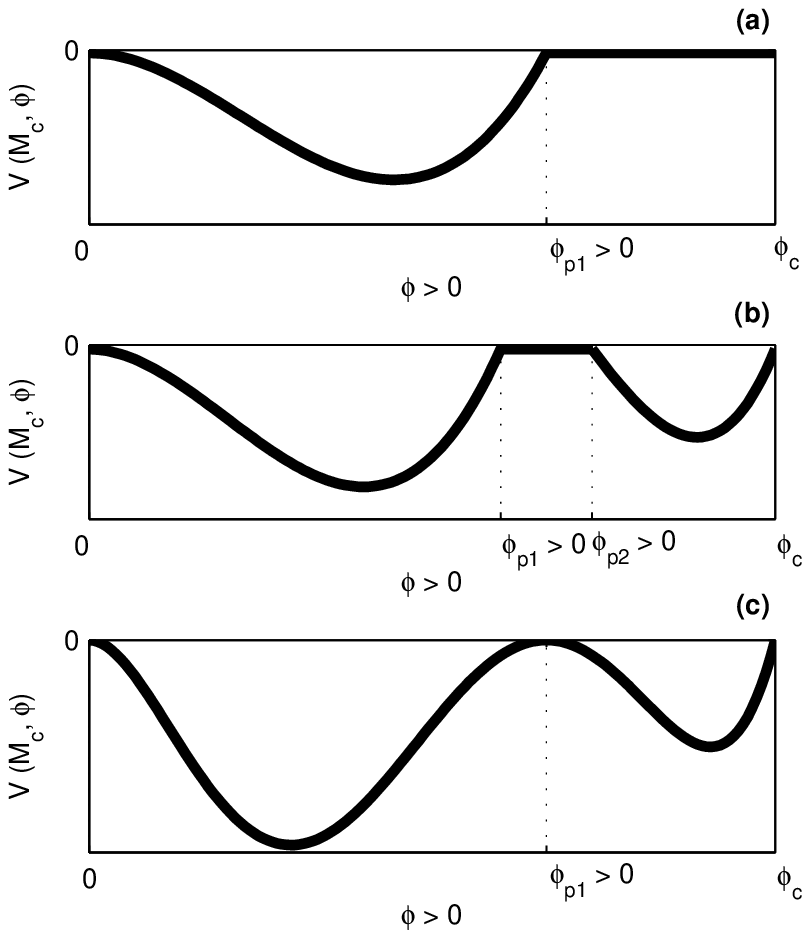}
  \caption{\label{fig:appendix} Figure (a) corresponds to the subcase 1 of case 2 of Theorem 6. Figure (b) corresponds to the subcase 2 of case 2 of Theorem 6. Figure (c) corresponds to the subcase 3 of case 2 of Theorem 6.}
\end{center}
\end{figure}

For \textbf{Case - 2}, there exists at least one $\phi \in (0,\phi_{c})$ such that $V(M_{c},\phi)= 0$. Let $\phi_{p1}$ be the smallest strictly positive root of the equation $V(M_{c},\phi)= 0$ such that $\phi_{p1} \in (0,\phi_{c})( \Leftrightarrow 0< \phi_{p1} < \phi_{c})$ and consequently, for this case, we have the following subcases.
\begin{description}
	\item[Subcase-1 ] $V(M_{c},\phi)<0$ for all $0<\phi<\phi_{p1}$ and $V(M_{c},\phi)=0$ for all $\phi \in [\phi_{p1} , \phi_{c}]$(See FIG. \ref{fig:appendix}(a)).
	\item[Subcase-2 ] $V(M_{c},\phi)<0$ for all $0<\phi<\phi_{p1}$ and $V(M_{c},\phi)=0$ for all $\phi \in [\phi_{p1} , \phi_{p2}]$ with $\phi_{p2}\leq \phi_{c}$(See FIG. \ref{fig:appendix}(b)).
	\item[Subcase-3 ] $V(M_{c},\phi)<0$ for all $0<\phi<\phi_{p1}$ and $V(M_{c},\phi)<0$ for all $\phi \in (\phi_{p1}-\epsilon_{p1},\phi_{p1})$ as well as $\phi \in (\phi_{p1},\phi_{p1}+\epsilon_{p1})$ along with $V(M_{c},\phi_{p1})=0$ for some $\epsilon_{p1}>0$ (See FIG. \ref{fig:appendix}(c)).
\end{description}

As $V(M_{c},\phi)$ is a continuously differentiable function of $\phi$ for all $\phi \in \Phi$ and for the \textbf{Subcase-1}, it is simple to check that $V(M_{c},\phi)$ is not differentiable at $\phi = \phi_{p1}$, \textbf{Subcase-1} is not possible. With the help of same argument, it is simple to check that \textbf{Subcase-2} is not possible when $\phi_{p1} < \phi_{p2}$. When $\phi_{p1} = \phi_{p2}$, then either we have \textbf{Subcase-3} or we have two different branches of the curve $V(M_{c},\phi)$ originated from $\phi = \phi_{p1}$. The later case is not possible because $V(M_{c},\phi)$ is not differentiable at $\phi = \phi_{p1}$. So, \textbf{Subcase-3} is the only possible case of \textbf{Case-2} and for this subcase we have the following conclusions.
 
\textbf{Subcase-3 :} In this case, as $V(M_{c},\phi)$ is a continuously differentiable function of $\phi$, $V(M_{c},\phi)$ is not only attains its maximum value at $\phi=\phi_{p1}$, $V(M_{c},\phi)$ is convex for all $\phi \in (\phi_{p1}-\epsilon_{p1},\phi_{p1}+\epsilon_{p1})$. Therefore, from the elementary theory of convexity of a function, we have 
\begin{eqnarray}\label{deldel2Mcphip1}
\frac{\partial V}{\partial \phi} \bigg|_{(M_{c},\phi_{p1})}=0 ~\mbox{   and   }~ \frac{\partial^{2} V}{\partial \phi^{2}} \bigg|_{(M_{c},\phi_{p1})}<0.
\end{eqnarray} 
Combining these two conditions along with the first condition of \textbf{Subcase-3}, i.e., $V(M_{c},\phi)<0$ for all $0<\phi<\phi_{p1}$, we get
\begin{description}
  \item[(1)] $V(M_{c},\phi_{p1})=0$,
	\item[(2)] $\frac{\partial V}{\partial \phi} \bigg|_{(M_{c},\phi_{p1})}=0$,
  \item[(3)] $\frac{\partial^{2} V}{\partial \phi^{2}}\bigg|_{(M_{c},\phi_{p1})}<0$,
	\item[(4)] $\phi_{p1}>0$,
	\item[(5)] $V(M_{c},\phi)<0$ for all $0<\phi<\phi_{p1}$.
\end{description}
These five conditions confirm the existence of PPDL solution of the energy integral (\ref{energy int}) at $M=M_{c}$. Therefore, under the conditions of the present theorem, one can get either a PPSW or a PPDL as a solution of the energy integral (\ref{energy int}) at $M=M_{c}$.\\
\textbf{Theorem 7 :} If $V(M,0)=V'(M,0)=V''(M_{c},0)=0$, $V'''(M_{c},0)>0$, $\partial V/\partial M<0$ for all $M>0$ and for all $\phi<0$, and if there exists at least one value $M_{0}$ of $M$ such that the system supports Negative Potential Solitary Waves (NPSWs) for all  $M_{c}<M< M_{0}$, then there exist either a NPSW or a Negative Potential Double Layer (NPDL) at $M=M_{c}$.\\
\textbf{Proof :} Same as Theorem 6 with a slight modification for the case of negative potential side.\\
\textbf{Theorem 8 :} If $V(M,0)=V'(M,0)=V''(M_{c},0)=0$, $V'''(M_{c},0)<0$, $\partial V/\partial M<0$ for all $M>0$ and for all $\phi>0$, and if the system supports only NPSWs for $M>M_{c}$, then there does not exist any PPSW at $M=M_{c}$.\\
\textbf{Proof :} From \textbf{Theorem 3}, we have seen that if $V(M,0)=V'(M,0)=V''(M_{c},0)=0$ and $V'''(M_{c},0)<0$, there may exist solitary wave in the positive potential side at $M=M_{c}$ but there does not exist any solitary wave in the negative potential side at $M=M_{c}$. If possible let there exists a PPSW at $M=M_{c}$ and for $M>M_{c}$, the system supports only NPSWs. Therefore, if $\phi_{c}$ is the amplitude of the PPSW at $M=M_{c}$, the following conditions are satisfied:
\begin{description}
  \item[Mc1:] $V(M_{c},\phi_{c})=0$,
	\item[Mc2:] $\phi_{c}>0$,
	\item[Mc3:] $\frac{\partial V}{\partial \phi}\bigg|_{(M_{c},\phi_{c})}>0$,
	\item[Mc4:] $V(M_{c},\phi)<0$ for all $0<\phi<\phi_{c}$.
\end{description}
As $\partial V/\partial \phi$ is a continuous function in its domain of definition, we have,
\begin{description}
  \item[Mc5:] $\frac{\partial V}{\partial \phi}$ is continuous in some rectangular neighborhood of $(M_{c},\phi_{c})$.
\end{description}

Therefore, all the assumptions, viz., \textbf{Mc1}, \textbf{Mc5} and \textbf{Mc3}, of \textbf{IFT} at $(M_{c},\phi_{c})$ are satisfied by the function $V(M,\phi)$. Therefore, from IFT, we can conclude that there exists a rectangular neighborhood $N(M_{c},\phi_{c};\delta_{1},\delta_{2})=N(M_{c},\delta_{1}) \times N(\phi_{c},\delta_{2})$ of $(M_{c},\phi_{c})$ such that corresponding to every $M \in N(M_{c},\delta_{1})$, $\phi (\in N(\phi_{c},\delta_{2}))$ can be expressed uniquely as a function of $M$, for which the following conditions are simultaneously satisfied.
\begin{description}
	\item[IFT Mc1 :] $\phi(M_{c})=\phi_{c}$,
	\item[IFT Mc2 :] $V(M,\phi(M))=0$ for all $M \in N(M_{c},\delta_{1})$,
	\item[IFT Mc3 :] $\frac{d \phi}{d M}=-(\frac{\partial V}{\partial M} \div \frac{\partial V}{\partial \phi})$ for all $(M, \phi) \in N(M_{c},\phi_{c};\delta_{1},\delta_{2})=N(M_{c},\delta_{1}) \times N(\phi_{c},\delta_{2})$.
\end{description}
Again from \textbf{Mc3}, using \textbf{BTh2}, we have two strictly positive real numbers $\delta_{3}(>0)$ and $\delta_{4}(>0)$ such that
\begin{description}
	\item[Mc6 :] $\frac{\partial V}{\partial \phi}>0$ for all $(M,\phi) \in N(M_{c},\phi_{c};\delta_{3},\delta_{4})=N(M_{c},\delta_{3}) \times N(\phi_{c},\delta_{4})$.
\end{description}
So, if we take $\delta=\min \{\delta_{1}, \delta_{3}\}$ and $\rho=\min \{\delta_{2}, \delta_{4}\}$, then $N(M_{c},\delta) \subseteq N(M_{c},\delta_{i})$ for all $i=1,3$, and $N(\phi_{c},\rho)\subseteq N(\phi_{c},\delta_{j})$ for all $j=2,4$, consequently, $N(M_{c},\phi_{c};\delta,\rho)=N(M_{c},\delta) \times N(\phi_{c},\rho) \subseteq N(M_{c},\delta_{i}) \times N(\phi_{c},\delta_{j}) = N(M_{c},\phi_{c};\delta_{i},\delta_{j})$ for all $i=1,3$ and $j=2,4$. Therefore, we can also apply IFT in the rectangular neighborhood  $N(M_{c},\phi_{c};\delta,\rho)=N(M_{c},\delta) \times N(\phi_{c},\rho)$, and consequently,
\textbf{IFT Mc1}, \textbf{IFT Mc2}, \textbf{IFT Mc3}, and \textbf{Mc6} can be re-written as follows:
\begin{description}
	\item[mMc1 :] $\phi(M_{c})=\phi_{c}$,
	\item[mMc2 :] $V(M,\phi(M))=0$ for all $M \in N(M_{c},\delta)$,
	\item[mMc3 :] $\frac{d \phi}{d M}=-(\frac{\partial V}{\partial M} \div \frac{\partial V}{\partial \phi})$ for all $(M,\phi) \in N(M_{c},\phi_{c};\delta,\rho)=N(M_{c},\delta) \times N(\phi_{c},\rho)$,
	\item[mMc6 :] $\frac{\partial V}{\partial \phi}>0$ for all $(M,\phi) \in N(M_{c},\phi_{c};\delta,\rho)=N(M_{c},\delta) \times N(\phi_{c},\rho)$.
\end{description}

Again, from the condition of the present theorem, we have $\partial V/\partial M<0$ for all $M>0$ and for all $\phi> 0$. As our discussion is restricted to positive potential side ($\phi >0$) only, from \textbf{mMc3} and \textbf{mMc6}, we get $\phi(M)$ is an increasing function of $M$ for all $M_{c}-\delta<M<M_{c}+\delta$. Consider a value $M_{1}$ of $M$ such that $M_{c}<M_{1}<M_{c}+\delta \Rightarrow M_{c}-\delta<M_{c}<M_{1}<M_{c}+\delta \Rightarrow  M_{c}-\delta<M_{1}<M_{c}+\delta \Rightarrow M_{1} \in N(M_{c},\delta)$, and consequently, from \textbf{mMc2}, the condition that the function $\phi(M)$ is an increasing function of $M$ for all $M_{c}-\delta<M<M_{c}+\delta$ and \textbf{mMc6}, we get the following three conditions:
\begin{description}
  \item[M1-1:] $V(M_{1},\phi_{1})=0$, with $\phi_{1}=\phi(M_{1})$,
	\item[M1-2:] $M_{c}-\delta<M_{c}<M_{1}<M_{c}+\delta \Rightarrow M_{c},M_{1} \in N(M_{c},\delta) \mbox{  with  } M_{c}<M_{1} \Rightarrow \phi(M_{c})<\phi(M_{1}) \Rightarrow \phi_{c}<\phi_{1} \Rightarrow \phi_{1}>\phi_{c}>0$,
	\item[M1-3:] $\frac{\partial V}{\partial \phi}\bigg|_{(M_{1},\phi_{1})}>0$,
\end{description}
[It is important to note that if we take $M \in (M_{c}-\delta,M_{c})$, then also the above three conditions are simultaneously satisfied but $M < M_{c} \Leftrightarrow V''(M,0)>0$, and consequently, $\phi=0$ is a position of stable equilibrium. So, there is no question of existence of any solitary wave solution of the energy integral (\ref{energy int}) for all $M \in (M_{c}-\delta,M_{c})$.]

Now if $V(M_{1},\phi)<0$ for all $0<\phi<\phi_{1}$, then from \textbf{M1-1}, \textbf{M1-2}, and \textbf{M1-3}, we can conclude that there exists a PPSW at $M=M_{1}>M_{c}$ but this is impossible as the system supports only NPSWs for $M>M_{c}$. Therefore, there exists at least one value $\phi_{2}$ of $\phi$ such that 
\begin{eqnarray}\label{VM1phi20}
V(M_{1},\phi_{2})\geq 0~\mbox{ for }~0<\phi_{2}<\phi_{1}.
\end{eqnarray}
Now we shall prove that actually $\phi_{2}$ is restricted by the following inequality: $\phi_{c}<\phi_{2}<\phi_{1}$.

Now from the condition $\partial V/\partial M<0$ for all $M>0$ and for all $\phi>0$, we have $V(M,\phi)$ is a strictly decreasing function of $M$ for all $M >0$ and for any given value of $\phi >0$. As $M_{1}>M_{c}$, we get
\begin{description}
  \item[M1-4 :] $V(M_{1},\phi)<V(M_{c},\phi)$ for all $\phi >0$.
\end{description}
Again, from \textbf{Mc4}, we get $V(M_{c},\phi)<0$ for all $0 <\phi <\phi_{c}$ $\Rightarrow$ $V(M_{1},\phi)<V(M_{c},\phi)<0$ for all $0 <\phi <\phi_{c}$, where we have used the inequality ($V(M_{1},\phi)<V(M_{c},\phi)$ for all $\phi >0$) as given in \textbf{M1-4}. Therefore, we have
\begin{description}
  \item[M1-5 :] $V(M_{1},\phi)<0$ for all $0<\phi < \phi_{c}$.
\end{description}

Again, as $\phi_{c}>0$ and $V(M_{c},\phi_{c})=0$, putting $\phi=\phi_{c}$ in \textbf{M1-4}, we get
\begin{description}
  \item[M1-6 :] $V(M_{1},\phi_{c})<V(M_{c},\phi_{c})=0 \Rightarrow V(M_{1},\phi_{c})<0$.
\end{description}
Combining \textbf{M1-5} and \textbf{M1-6}, we get
\begin{description}
  \item[M1-7 :] $V(M_{1},\phi)<0$ for all $0<\phi \leq \phi_{c}$.
\end{description}

Therefore, from \textbf{M1-7}, we can conclude that $\phi_{c}<\phi_{2}<\phi_{1}$. Now we shall prove that there exists a value $M_{2}$ of $M$ such that $\phi_{2}=\phi(M_{2})$ for $M_{c}<M_{2}<M_{1}$.

Let us define the function $\psi(M)=\phi(M)-\phi_{2}$ for all $M \in [M_{c}, M_{1}]$. Now as $\phi(M)$ is a continuously differentiable function of $M$ for all $M \in N(M_{c},\delta)$, then it is automatically continuous for all $M \in N(M_{c},\delta)$. Again as $[M_{c}, M_{1}] \subset N(M_{c},\delta)$, $\phi(M)$ is also continuous for all $M \in [M_{c},M_{1}]$, and consequently, $\psi(M)$ is continuous for all $M \in [M_{c},M_{1}]$. Now as $\phi_{c}=\phi(M_{c})$ and $\phi_{1}=\phi(M_{1})$, the condition $\phi_{c}<\phi_{2}<\phi_{1}$ gives $\phi(M_{c})<\phi_{2}<\phi(M_{1}) $ $\Leftrightarrow $ $\phi(M_{c})-\phi_{2}<0$ and $\phi(M_{1})-\phi_{2}>0$ $\Leftrightarrow $ $\psi(M_{c})<0$ and $\psi(M_{1})>0$.

Therefore, we have a continuous function $\psi(M):[M_{c},M_{1}]\longmapsto \Re$ such that $\psi(M_{c})<0$ and $\psi(M_{1})>0$,and consequently, by \textbf{Bolzano's theorem}, we can conclude that there exists a value $M_{2}$ of $M$ such that $\psi(M_{2})=0 \Leftrightarrow \phi(M_{2})=\phi_{2}$ for $M_{c}<M_{2}<M_{1}$. Now as $V(M,\phi)$ is strictly decreasing function of $M >0$ for any fixed value of $\phi>0$, and consequently, as $M_{2}<M_{1}$, we have $V(M_{2},\phi)>V(M_{1},\phi)$ for any given value of $\phi > 0$. If we take $\phi=\phi_{2}$, we get $V(M_{2},\phi_{2})>V(M_{1},\phi_{2})\Leftrightarrow V(M_{1},\phi_{2}) < V(M_{2},\phi_{2})=V(M_{2},\phi(M_{2}))=0$ because $M_{c}<M_{2}<M_{1}<M_{c}+\delta \Rightarrow M_{2} \in N(M_{c},\delta) \Rightarrow V(M_{2},\phi(M_{2}))=0$, where we have used the condition \textbf{mMc2} to get $V(M_{2},\phi(M_{2}))=0$, So, we have $V(M_{1},\phi_{2}) < 0$ which again contradicts Eq. (\ref{VM1phi20}), i.e., $V(M_{1},\phi_{2}) \geq 0$. Thus, our original assumption is wrong, and consequently, theorem is proved, i.e., if the system supports only NPSWs for $M>M_{c}$, then there does not exist any PPSW at $M=M_{c}$ provided the conditions of the present theorem are fulfilled.\\
\textbf{Theorem 9 :} If $V(M,0)=V'(M,0)=V''(M_{c},0)=0$, $V'''(M_{c},0)>0$, $\partial V/\partial M<0$ for all $M>0$ and for all $\phi<0$, and if the system supports only PPSWs for $M>M_{c}$, then there does not exist any NPSW at $M=M_{c}$.\\
\textbf{Proof :} Same as Theorem 8 with a slight modification for the negative potential side.\\
\textbf{Theorem 10 :} If $V(M,0)=V'(M,0)=V''(M_{c},0)=0$, $V'''(M_{c},0)\neq 0$, $\partial V/\partial M<0$ for all $M>0$ and for all $\phi \neq 0$, and if there exists at least one value $M_{0}$ of $M$ such that the system supports both positive and NPSWs for all  $M_{c}<M< M_{0}$, i.e., if the system supports the coexistence of both PPSWs and NPSWs for all $M_{c}<M< M_{0}$, then if $V'''(M_{c},0)<0$, there exist either a PPSW or a PPDL at $M=M_{c}$ whereas if $V'''(M_{c},0)>0$ there exist either a NPSW or a NPDL at $M=M_{c}$.\\
\textbf{Proof :} Follows from \textbf{Theorem 6} and \textbf{Theorem 7}.

From theorem 6 -theorem 10, it is simple to see that the critical Mach number alone cannot determine the existence of particular nonlinear structure and its polarity. This is not even the motivation of our present manuscript. However, our intension is to determine the conditions under which nonlinear structure exists at the critical Mach number. The existence of nonlinear structure at the critical Mach number and its polarity is determined by the nature of the solitary structures for at least one right neighbourhood of the point $M=M_{c}$ together with the sign of the function $V'''(M_{c}, 0)$.

Except Theorem 1 and Theorem 2, the proof of rest of the theorems (Theorem 3 - Theorem 10) carry physical sense inherently, since the proofs are based on the physical interpretation of the energy integral as discussed in Sec. II. For example, Theorem 6 is proved with the help of Theorem 3 while Theorem 3 is proved on the basis of physical interpretation of the energy integral. One can get new result behind the nonlinear structures at $M=M_{c}$ by considering any specific problem, only when these theorems have been used on the qualitatively different solution spaces or compositional parameter spaces with respect to any parameter of the system (keeping arbitrary fixed values of other parameters in their respective physically admissible range) showing the existence region of different types of solitary structures for $M>M_{c}$. For this purpose, we have considered the same problem of dust acoustic wave in nonthermal plasma of Das \textsl{et al.}\cite{das09} in Sec. \ref{sec:verify} 

So, without going through all qualitatively different solution spaces of the energy integral by considering the entire range of parameters involved in the system, it is not possible to make a systematic investigation of the solitary structures at $M=M_{c}$. To construct the solution space or compositional parameter space with respect to any parameter of the system (keeping arbitrary fixed values of other parameters in their respective physically admissible range) showing the existence region of different types of solitary structures for $M>M_{c}$, it is not necessary to know the existence of solitary structure at $M=M_{c}$. For $M>M_{c}$, using the physical interpretation as given in first para of Sec. II, one can easily set up a numerical scheme to construct the compositional parameter space with respect to any parameter of the system (keeping arbitrary fixed values of other parameters in their respective physically admissible range) showing the existence region of different types of solitary structures. So, to construct the solution space, the existence of solitary structure at $M=M_{c}$ does not play any role.
\section{\label{sec:verify}Application}
Here we consider the specific problem of Das \textsl{et al}\cite{das09} to apply the analytical results (theorem 3 - theorem 10) as presented in this paper. In this paper,\cite{das09} a computational scheme has been developed to study the arbitrary amplitude dust acoustic solitary waves and double layers in a nonthermal plasma consisting of negatively charged dust grains, nonthermal ions and isothermal electrons including the effect of dust temperature. The Sagdeev potential approach, which is valid to study the arbitrary amplitude solitary waves and double layers, has been employed. Four basic parameters of the system are $\mu, \alpha, \beta_{1}$ and $\sigma_{d}$, which are respectively the ratio of unperturbed number density of electrons to that of nonthermal ions, the ratio of average temperature of nonthermal ions to that of isothermal electrons, a parameter of the nonthermal distribution of ions and the ratio of average temperature of dust particles to that of ions divided by the number of negative charges residing on a dust grain surface. The Sagdeev potential $V(\phi)(\equiv V(M,\phi))$ as given by Eq. (21) of Das \textsl{et al.},\cite{das09} is well defined as a real number for all $M \in \mathcal{M}$ and for all $\phi \in \Phi$, where $\mathcal{M}$ is the set of all stictly positive real number and $\Phi = \{\phi : \Psi_{M} \leq \phi < +\infty \}$, and $\Psi_{M}$ is given by the first equation of (17) of Das \textsl{et al.}\cite{das09} So, except the point $\phi = \Psi_{M}$, any $\phi \in \Phi$ is an interior point of $\Phi$ and any $M>0$ is an interior point of $\mathcal{M}$. The solution spaces of the energy integral (20) with respect to $\beta_{1}$ has been shown in FIG. 3 of Das \textsl{et al.}\cite{das09} for fixed values of other parameters. From this solution space, they have found that for any fixed values of the parameters $\mu$, $\alpha$ and $\sigma_{d}$ the entire interval of $\beta_{1}$ can be broken up into four disjoint subintervals I : $0 \leq \beta_{1} < \beta_{1c}$, II : $\beta_{1c} \leq \beta_{1} \leq \beta_{2c}$, III : $\beta_{2c} < \beta_{1} \leq \beta_{3c}$, IV : $\beta_{3c} < \beta_{1} < \beta_{M}$, where $\beta_{M}=\min\{1+\alpha\mu,~4/3\}$. FIG. 3 of Das \textsl{et al.}\cite{das09} also shows that (i) in subinterval I, only NPSWs can exist and the Mach number $M$ for these waves lies within the interval $M_{c} < M \leq M_{max}$, where $M_{c}$ is the lower bound of $M$ and $M_{max}$ is the upper bound of $M$ which is defined only when the system can support NPSWs, (ii) in subinterval II, both NPSWs and PPSWs can coexist and the Mach number $M$ for these waves lies within the interval $M_{c} < M < M _{D}$, whereas only NPSWs are possible if $M_{D} < M  \leq M _{max}$, (iii) in subinterval III, both NPSWs and PPSWs can coexist and Mach number for these waves lies within $M_{c} < M \leq M_{max}$, whereas only PPSWs are possible if $M_{max} < M < M_{D}$, (iv) in subinterval IV, only PPSWs can exist and Mach number $M$ for these waves lies within $M_{c} < M < M_{D}$. In subintervals II, III and IV, only PPDL solution is possible along the curve $M=M_{D}$. Again from this figure, we see that the curve $M=M_{D}$ tends to intersect the curve $M=M_{c}$ at $\beta_{1}=\beta_{1c}$, i.e., there always exists a PPDL solution in any right neighborhood of $M_{c}$ for some $\beta_{1}>\beta_{1c}$. Therefore, at $\beta_{1} = \beta_{1c}$, we can expect a PPDL solution at $M=M_{c}$, if the other conditions of Theorem 6 have been fulfilled. Before going to discuss the existence of solitary wave and/or double layer solutions of the present problem at $M=M_{c}$, it is necessary to know the sign of $\partial V/\partial M$, $V'''(M_{c},0)$ and the sign of $V''''(M_{c},0)$ when $V'''(M_{c},0)=0$. From the expression of $V(\phi)(\equiv V(M,\phi))$ as given by Eq. (21) of Das \textsl{et al.},\cite{das09} we get the following equations. 
\begin{eqnarray}\label{delvdelM}
\frac{\partial V}{\partial M} = -M\bigg(\sqrt{n_{d}}-\frac{1}{\sqrt{n_{d}}}\bigg)^{2},
\end{eqnarray}
\begin{equation}\label{v3 at Mc non}
    V'''(M_{c}, 0) = -\frac{12 \sigma_{d} \beta_{T}^{3} + 3 \mu_{T}\beta_{T}^{2} -\mu_{T}^{2}(1-\alpha^{2}\mu)}{\mu_{T}^{3}},
\end{equation}
where
\begin{equation}
    \beta_{T} = 1+\alpha \mu-\beta_{1}, \mu_{T} = 1- \mu.
\end{equation}
From Eq. (\ref{delvdelM}), we see that $\partial V/\partial M<0$ for all $M>0$ and for all $\phi\neq 0$, and consequently, for all $M>0$, the condition $\partial V/\partial M<0$ is satisfied for all $\phi > 0$ as well as $\phi < 0$.

Now from the expression of $V'''(M_{c},0)$ as given by Eq. (\ref{v3 at Mc non}) of present paper, it can be easily checked that $(\partial/\partial \beta_{1}) (V'''(M_{c}, 0))>0$ for all $0 \leq \beta_{1}< \beta_{M}$ and for any admissible values of the other parameters, consequently, $V'''(M_{c}, 0)$ is strictly increasing function of $\beta_{1}$ for all $0 \leq \beta_{1}< \beta_{M}$. Again, it can be easily checked that $V'''(M_{c}, 0)<0$ for $\beta_{1}=0$ and $V'''(M_{c}, 0)>0$ for $\beta_{1}=1+\alpha\mu \geq \beta_{M}$. Therefore, $V'''(M_{c}, 0)$ strictly increases with increasing $\beta_{1}$ starting from a negative value and ending with a positive value. So, $V'''(M_{c}, 0)$ intersects the axis of $\beta_{1}$ at $\beta_{1}=\beta_{c}(0 < \beta_{c}<\beta_{M})$ and consequently, we have $V'''(M_{c}, 0)<0$ for all $0 \leq \beta_{1}<\beta_{c}$, $V'''(M_{c}, 0)>0$ for all $\beta_{c} < \beta_{1}<\beta_{M}$ and $V'''(M_{c}, 0)=0$ at $\beta_{1} = \beta_{c}$. It is also interesting to note that the equation $V'''(M_{c}, 0)=0$ has one and only one real root $\beta_{1} = \beta_{c}$ of $\beta_{1}$ within the admissible range of $\beta_{1}$ and for any  fixed values of the other parameters. Again, it is a simple task to check that $V''''(M_{c}, 0)>0$ at $\beta_{1} = \beta_{c}$, i.e., $V''''(M_{c}, 0)>0$ when $V'''(M_{c}, 0)=0$. Therefore, when $V'''(M_{c}, 0)=0$, $\phi=0$ is the position of stable equilibrium and consequently, there does not exist any solitary structure of the energy integral (20) of Das \textsl{et al.},\cite{das09} at $M=M_{c}$. Now we are in a position to discuss the existence of solitary wave and/or double layer solution of the energy integral (20) of Das \textsl{et al.},\cite{das09} at $M=M_{c}$ for any admissible value of $\beta_{1}$ and fixed values of the other parameters. 

In subinterval I ($0 \leq \beta_{1} < \beta_{1c}$), one can expect only NPSW at $M=M_{c}$, provided $V'''(M_{c}, 0) > 0$ (from Theorem 4). But it can be easily checked that $V'''(M_{c}, 0) < 0$ for any value of $\beta_{1}$ lying within subinterval I. So from Theorem 8, we can conclude that there does not exist any solitary wave solution at $M=M_{c}$ for any value of $\beta_{1}$ lying within subinterval I. Again, in subinterval IV ($\beta_{3c} < \beta_{1} < \beta_{M}$), one can expect only PPSW at $M=M_{c}$, provided $V'''(M_{c}, 0) < 0$ (from Theorem 3). But it can be easily checked that $V'''(M_{c}, 0) > 0$ for any value of $\beta_{1}$ lying within subinterval IV. So from Theorem 9, we can conclude that there does not exist any solitary wave solution at $M=M_{c}$ for any value of $\beta_{1}$ lying within subinterval IV. As the system supports coexistence of both NPSWs and PPSWs in subintervals II and III, i.e., for $\beta_{1}$ lying within the interval $\beta_{1c} \leq \beta_{1} \leq \beta_{3c}$, from Theorem 10, we can conclude that there must exist solitary wave or double layer solution at $M=M_{c}$ when $\beta_{1c} \leq \beta_{1} \leq \beta_{3c}$ and the nature of the solitary wave or double layer solution depends on the sign of $V'''(M_{c}, 0)$.

\begin{figure}
\begin{center}
  \includegraphics{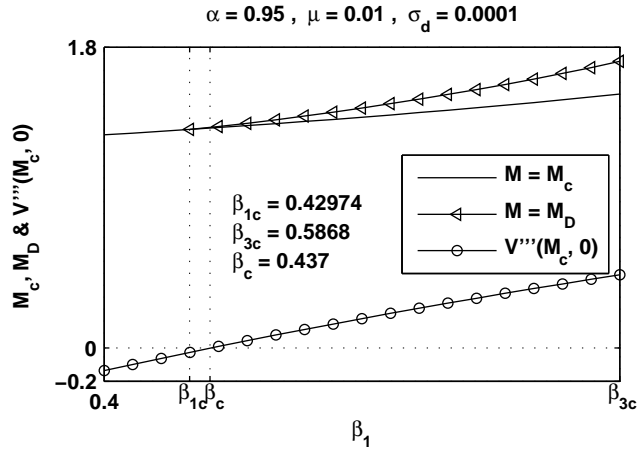}
  \caption{\label{fig:verify} $M_{c}$, $M_{D}$ and $V'''(M_{c}, 0)$ are plotted against $\beta_{1}$ for values of $\beta_{1}$ lies within the intervals $\beta_{1c}\leq \beta_{1}\leq \beta_{3c}$. At $\beta_{1} = \beta_{1c}$, the curve $M=M_{D}$ tends to intersect the curve $M=M_{c}$.}
\end{center}
\end{figure}

In FIG. \ref{fig:verify}, $M_{c}$, $M_{D}$ and $V'''(M_{c}, 0)$ have been plotted against $\beta_{1}$ for $\beta_{1c} \leq \beta_{1} \leq \beta_{3c}$. From this figure we see that the double layer solution is not possible for $\beta_{1c} < \beta_{1} \leq \beta_{3c}$, as the curve $M=M_{D}$ does not tend to intersect the curve $M=M_{c}$ for any point of $\beta_{1}$ lying within the interval $\beta_{1c} < \beta_{1} \leq \beta_{3c}$. So, for any $\beta_{1}$ lying within the interval $\beta_{1c} < \beta_{1} \leq \beta_{3c}$, we get either a NPSW or a PPSW at $M=M_{c}$. Again from FIG. \ref{fig:verify}, we see that the curve $M=M_{D}$ tends to intersect the curve $M=M_{c}$ at $\beta_{1}=\beta_{1c}$, i.e., there always exists a PPDL solution in any right neighborhood of the point $M_{c}$ for some $\beta_{1}>\beta_{1c}$. Therefore, at $\beta_{1} = \beta_{1c}$, we have a PPDL solution at $M=M_{c}$. The existence of solitary wave solution at $M=M_{c}$ for any $\beta_{1}$ lying within the interval $\beta_{1c} < \beta_{1} \leq \beta_{3c}$ and the existence of PPDL solution at $M=M_{c}$ when $\beta_{1} = \beta_{1c}$ can easily be verified through FIG. \ref{fig:verify}. Specifically, we have $V'''(M_{c}, 0) < 0$ for any $\beta_{1}$ lying within the interval $\beta_{1c}<\beta_{1}<\beta_{c}$ and $V'''(M_{c}, 0) > 0$ for any $\beta_{1}$ lying within the interval $\beta_{c}<\beta_{1}\leq\beta_{3c}$. Consequently, we have PPSW at $M=M_{c}$ for any $\beta_{1}$ lying within the interval $\beta_{1c}<\beta_{1}<\beta_{c}$, whereas we get NPSW at $M=M_{c}$ for any $\beta_{1}$ lying within the interval $\beta_{c}<\beta_{1}\leq\beta_{3c}$ and at $\beta_{1} = \beta_{c}$, there does not exist any solitary structure at $M=M_{c}$.

\begin{figure}
\begin{center}
  \includegraphics{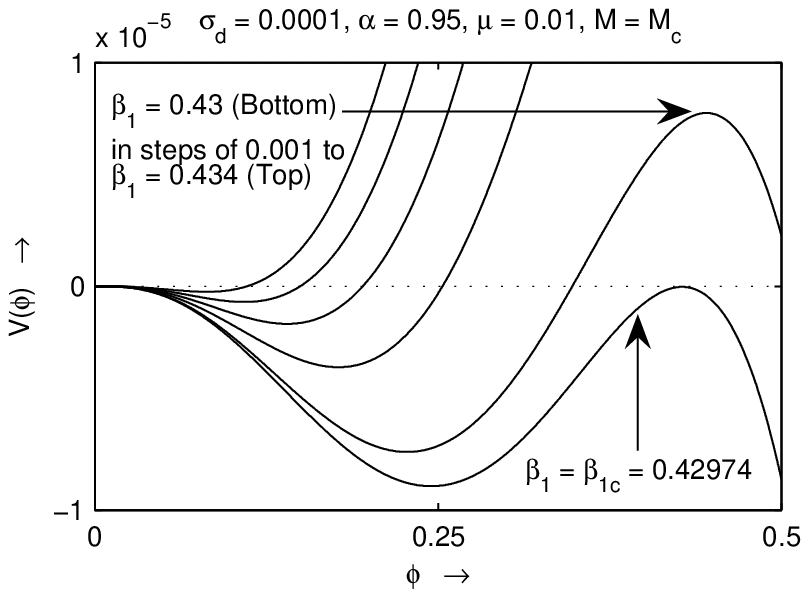}
  \caption{$V(\phi)$ is plotted against $\phi$ for different values of $\beta_{1}$ lies within the intervals $\beta_{1c}\leq \beta_{1}<\beta_{c}$. \label{fig:positive} }
\end{center}
\end{figure}

To be more specific, we draw FIG. \ref{fig:positive}. This figure shows that the amplitude of the PPSW increases with decreasing $M_{c}$ and ultimately this sequence of PPSWs end with a PPDL solution at $M=M_{c}$ when $\beta_{1}$ assumes the value $\beta_{1c}$. However, we have seen in the literature that when all the parameters involved in the system assume fixed values in their respective physically admissible range, the amplitude of solitary wave increases with increasing $M$ and these solitary waves end with a double layer of same polarity, if exists. Here it is important to note that $M$ is not a function of the parameters involved in the system but is restricted by the inequality $M_{c}<M<M_{D}$, where $M=M_{D}$ corresponds to a double layer solution. So we cannot compare this case with the case of $M_{c}$, since $M_{c}$ is a function of the parameters involved in the system and consequently, monotonicity of $M_{c}$ entirely depends on a parameter when the other parameters assume fixed values in their respective physically admissible range. But the solitons and double layer are not two distinct nonlinear structures even when $M=M_{c}$. Actually, double layer solution, if exists, must be the limiting structure of at least one sequence of solitons of same polarity even when $M=M_{c}$. More specifically, existence of double layer solution implies that there must exists a sequence of solitary waves of same polarity having monotonically increasing amplitude converging to the double layer solution, i.e., the amplitude of the double layer solution acts as an exact upper bound or Least Upper Bound (\textit{lub}) of the amplitudes of the sequence of solitary waves. Therefore, if the double layer solution exists at $M=M_{c}$ then this double layer solution is also a limiting structure of a sequence of solitary waves of same polarity. So, double layer solution can be regarded as a source or sink of solitary waves of same polarity for increasing Mach number. Particularly, for the problem of Das \textit{et al.} \cite{das09}, the double layer solution at $M=M_{c}$ when $\beta_{1} = \beta_{1c}$ can be regarded as a source for PPSWs for increasing $M_{c}$ along the curve $M=M_{c}$ whereas the same double layer solution at $M=M_{c}$ when $\beta_{1} = \beta_{1c}$ can be regarded as a sink of PPSWs for decreasing $M_{c}$ along the curve $M=M_{c}$. However, for any other double layer solution, i.e., for the double layer solution at $M=M_{D}(>M_{c})$ can be regarded as sink of PPSWs for increasing $M$ restricted by the inequality: $M_{c}<M<M_{D}$.

\begin{figure}
\begin{center}
  \includegraphics{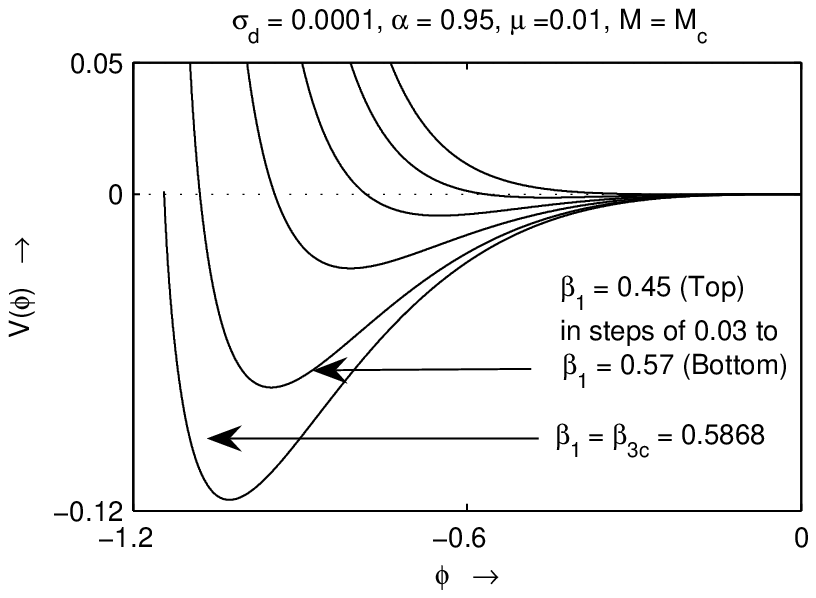}
  \caption{$V(\phi)$ is plotted against $\phi$ for different values of $\beta_{1}$ lies within the intervals $\beta_{c}<\beta_{1}\leq \beta_{3c}$.\label{fig:negative} }
\end{center}
\end{figure}

On the other hand, if we take the interval $\beta_{c}< \beta_{1} \leq \beta_{3c}$ and move along the curve $M=M_{c}$, then we have seen from FIG. \ref{fig:negative} that the amplitude of NPSW increases with increasing $M_{c}$ and it attains its maximum value at $\beta_{1} = \beta_{3c}$. This is, of course, expected since $\beta_{3c}$ is the point of intersection of $M_{c}$ and $M_{max}$ and for $\beta_{1} > \beta_{3c}$ there does not exist any solitary structure at $M=M_{c}$. For this interval, i.e., $\beta_{c}< \beta_{1} \leq \beta_{3c}$, the solitary wave solution at $M=M_{c}$ is similar to the solitary wave solution for $M>M_{c}$, because for this interval the amplitude of NPSW increases with increasing $M_{c}$.

\begin{figure}
\begin{center}
  \includegraphics{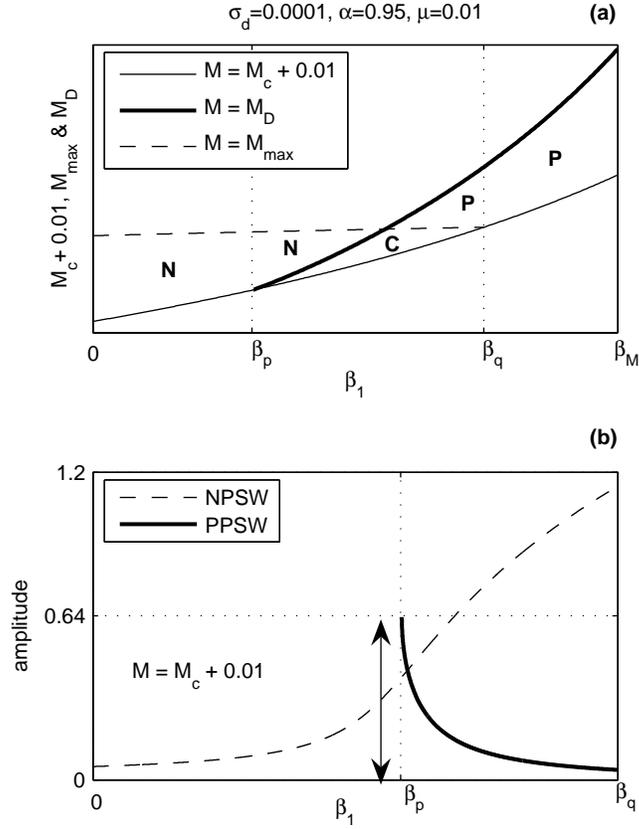}
  \caption{(a)solution space for the present system with respect to $\beta_{1}$. Here the curve $M=M_{c}$ is omitted from the solution space as presented in FIG. 3 of Ref [26]. (b) Variation in amplitude of NPSWs and PPSWs along the curve $M=M_{c}+0.01$. \label{fig:sol space amp}}
\end{center}
\end{figure}

Moreover, it can be easily checked that the solitary structures along the curve $M=M_{c}$ is same as the solitary structures for $M>M_{c}$, if one can move the solution space through the family of curves parallel to the curve $M=M_{c}$. To be more specific, solution space for the present system with respect to $\beta_{1}$ has been presented in FIG. \ref{fig:sol space amp}(a), in which the curve $M=M_{c}$ is omitted from the solution space as presented in FIG. 3 of Ref [26]. Now consider the family of curves parallel to $M=M_{c}$. For instance, consider one such parallel curve for $M=M_{c}+0.01$ as shown in FIG. \ref{fig:sol space amp}(a). In this figure, $\beta_{p}$ is the value of $\beta_{1}$ where the curve $M=M_{D}$ intersects the curve $M=M_{c}+0.01$, whereas $\beta_{q}$ is the value of $\beta_{1}$ where the curve $M=M_{max}$  intersects the curve $M=M_{c}+0.01$. FIG. \ref{fig:sol space amp}(a) can be interpreted in the same way of FIG. 3 of Ref [26] with $\beta_{1c}$ replaced by $\beta_{p}$ and $\beta_{3c}$ replaced by $\beta_{q}$. However, in FIG. \ref{fig:sol space amp}(a) the solitary structures exist along the curve $M=M_{c}+0.01$, specifically, (i) only NPSW exists for $0\leq\beta_{1}<\beta_{p}$, (ii) both NPSW and PPSW coexist for $\beta_{p}<\beta_{1}\leq\beta_{q}$, (iii) only PPSW exists whenever $\beta_{q}<\beta_{1}<\beta_{M}$ and at $\beta_{1}=\beta_{p}$, a NPSW coexists with a PPDL. The variation of amplitude of these solitary waves (exist along the curve $M=M_{c}+0.01$) have been shown in FIG. \ref{fig:sol space amp}(b). This figure shows that the amplitude of NPSWs increase with $\beta_{1}$ having maximum amplitude at $\beta_{1}=\beta_{q}$. Again, the amplitude of PPSWs decreases with increasing $\beta_{1}$ for $\beta_{1}>\beta_{p}$ and the amplitudes of PPSWs are bounded by the amplitude of PPDL at $\beta_{1}=\beta_{p}$. Moreover, the amplitudes of the PPDL solutions increases with increasing $M_{D}$ along the curve $M=M_{D}$. FIG. \ref{fig:sol space amp}(b) also shows that along the curve $M=M_{c}+0.01$, the amplitude of PPSW increases with decreasing $M$ along the curve $M=M_{c}+0.01$ and ultimately, these PPSWs end with a PPDL at $\beta_{1}=\beta_{p}$. This fact is similar for the case, when we move along the curve $M=M_{c}$. So, if we move the solution space along the family of curves parallel to the curve $M=M_{c}$, then one cannot distinguish between the solitary structures, exist along the curve $M=M_{c}$ and any curve parallel to $M=M_{c}$. For $\sigma_{d} = 0.0001$, $\alpha=0.95$ and $\mu=0.01$, the values of $\beta_{p}$ and $\beta_{q}$ lie in the neighborhood of $\beta_{1}=0.44475$ and $\beta_{1}=0.5809$, respectively. To form a PPDL, the positive energy must dominate the negative energy in some region of the system and the PPDL will occur at some point where the potential drop is maximum. From FIG. \ref{fig:sol space amp}(b), it is clear that the positive potential dominates the negative potential in a right neighborhood of $\beta_{1}=\beta_{p}$ and again the potential drop is maximum thereat. This figure shows that physical reasons also suggest the existence of PPDL solution at $\beta_{1}=\beta_{p}$, which is already confirmed in FIG. \ref{fig:sol space amp}(a). This is the reason for the coexistence of NPSW and PPDL at $\beta_{1}=\beta_{p}$. Actually, the same phenomena happens for the occurrence of PPDL solution at $M=M_{c}$, although there does not exist any NPSW along the curve $M=M_{c}$ for $0\leq \beta_{1}<\beta_{1c}$. We'll see later, that weakly nonlinear negative potential solitary structure exist along the curve $M=M_{c}$ for $0\leq \beta_{1}<\beta_{1c}$ and also the amplitude of the NPSW tends to zero as $M \rightarrow M_{c}$, whereas PPSW of finite amplitude exist along the curve $M=M_{c}$ in the right neighborhood of $\beta_{1}=\beta_{1c}$ and consequently, the potential drop is maximum at $\beta_{1}=\beta_{1c}$. This is the reason behind the occurrence of PPDL at $M=M_{c}$ when $\beta_{1}=\beta_{1c}$. This can also be interpreted as follows: Since there is no NPSW solution along the curve $M=M_{c}$ for $0\leq \beta_{1}<\beta_{1c}$, one can simply think the occurrence of NPSW of amplitude zero along the curve $M=M_{c}$ for $0\leq \beta_{1}<\beta_{1c}$ and in this case also the potential drop is maximum at $\beta_{1}=\beta_{1c}$ and consequently, we have a PPDL solution at $M=M_{c}$. So the solitary structures at $M=M_{c}$ is not localized. Actually, the solitary structure along the curve $M=M_{c}$ is same as the solitary structures for $M>M_{c}$, if we move the solution space through the family of curves parallel to the curve $M=M_{c}$.

Again, at $\beta_{1} = \beta_{c}$, $V'''(M_{c}, 0)=0$ and $V''''(M_{c}, 0)>0$, and consequently, $\phi=0$ is the position of stable equilibrium. Therefore, it is not possible to get any solitary structure of the energy integral (20) of Das \textsl{et al.},\cite{das09} at $M=M_{c}$ when $\beta_{1} = \beta_{c}$. From FIG. \ref{fig:positive}, we see that along the curve $M=M_{c}$ the amplitude of PPSW decreases with increasing $\beta_{1}$ for $\beta_{1c}<\beta_{1}<\beta_{c}$ and ultimately it collapses at $\beta_{1} = \beta_{c}$. On the other hand, from FIG. \ref{fig:negative}, we see that along the curve $M=M_{c}$ the amplitude of NPSW decreases with decreasing $\beta_{1}$ for $\beta_{c}<\beta_{1}\leq\beta_{3c}$ and ultimately it also collapses at $\beta_{1} = \beta_{c}$. So, the figures FIG. \ref{fig:positive} and FIG. \ref{fig:negative} are consistent with the physical interpretation for nonexistence of solitary structure at $M=M_{c}$ when $\beta_{1} = \beta_{c}$. Again, PPSWs and NPSWs at $M=M_{c}$ behave KdV like solitons in the left and right neighbourhood of the point $\beta_{1} = \beta_{c}$, respectively.

\begin{figure}
\begin{center}
  \includegraphics{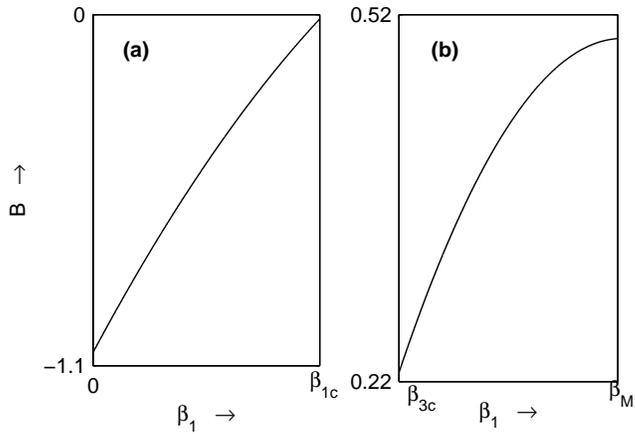}
  \caption{$B$ is plotted against $\beta_{1}$ for (a) $0 \leq \beta_{1}<\beta_{1c}$, (b) $\beta_{3c}< \beta_{1}<\beta_{M}$. \label{fig:B} }
\end{center}
\end{figure}

Again, the existence of solitary structures at $M=M_{c}$ indicates the applicability of small amplitude theory to study weakly nonlinear solitary structures. To be more specific, if we replace $V(\phi)$ by $V_{3}(\phi)$ to study weakly nonlinear solitary structures on the basis of the assumption $|\phi|\ll 1$, the energy integral (20) of Das \textsl{et al.} \cite{das09} assumes the following form:
\begin{eqnarray}\label{energy int3}
\frac{1}{2} \bigg(\frac{d\phi}{d\xi}\bigg)^{2}+ V_{3}(\phi) = 0,
\end{eqnarray}
where $V_{3}(\phi)$ is given by
\begin{eqnarray}
V_{3}(\phi) &=& \frac{1}{2} A \phi^{2} + \frac{1}{3} B \phi^{3}, \label{v3}
\end{eqnarray}
\begin{eqnarray}
A  &=& \frac{1}{M^{2} - 3 \sigma_{d}} -\frac{1-\beta_{1} + \alpha \mu}{1
- \mu}, \label{A}
\end{eqnarray}
\begin{eqnarray}
B &=& \frac{1}{2}\bigg[\frac{1-\alpha^{2} \mu}{1 - \mu} - \frac{3
(M^{2} + \sigma_{d})}{(M^{2} - 3 \sigma_{d})^{3}}\bigg].\label{B}
\end{eqnarray}
We do not consider the case $B = 0$, since the expression for $d^{2}\phi/d \xi^{2}$, in this case, does not contain any nonlinear term and therefore no solitary structures exist. For $B \neq 0$, we can write $V_{3}(\phi)$ as 
\begin{eqnarray}
V_{3}(\phi) = \frac{B}{3} \phi^{2}(\phi-\phi_{m}) = -\frac{B}{3} \phi^{2}(\phi_{m}-\phi) , \label{mv3}
\end{eqnarray}
where
\begin{eqnarray}
\phi_{m} =  -\frac{3A}{2B}, \label{phim}
\end{eqnarray}
Now applying the conditions for existence of the solitary structures of the energy integral (\ref{energy int3}) as discussed in first para of Sec. II, we have found that weakly nonlinear DAS wave exists if and only if $A < 0$ and $B \neq 0$. The condition $A < 0$ gives
\begin{eqnarray}\label{mgtmc}
  M>M_{c}=\sqrt{3 \sigma_{d} + \frac{1-\mu}{1+\alpha \mu - \beta_{1}}}.
\end{eqnarray}
So, $M_{c}$ is the lower bound of $M$ from which solitary wave solution of (\ref{energy int3}) starts to exist. The condition $B \neq 0$ determines the nature of the solitary wave solution of (\ref{energy int3}). In particular, the solitary wave solution of (\ref{energy int3}) describes either a PPSW or a NPSW according to $B>0$ or $B<0$. The amplitude of PPSW or NPSW is given by $|\phi_{m}|=-(3A)/(2|B|) \rightarrow 0$ as $M \rightarrow M_{c}+0$. Furthermore, solution of (\ref{energy int3}) is unable to describe the existence of both PPSW and NPSW simultaneously, and consequently, $V_{3}(\phi)$ is representative of $V(\phi)$ to study weakly nonlinear solitary structures only when $\beta_{1}$ lies either within the interval $0 \leq \beta_{1} < \beta_{1c}$ or within the interval $\beta_{3c} < \beta_{1} < \beta_{M}$ since the amplitude of the solitary wave solution of (\ref{energy int3}) tends to zero as $M \rightarrow M_{c}+0$. In FIG. \ref{fig:B}, $B$ is plotted against $\beta_{1}$ for $M = M_{c}+0.00001$. Although, from FIG. \ref{fig:B}(a), we see that $B$ is equal to zero at $\beta_{1}=\beta_{1c}$, one can easily verify that $B$ becomes zero in the neighborhood of $\beta_{1}=\beta_{c}$ when $M$ is close to $M_{c}$. Therefore FIG. \ref{fig:B}(a) and FIG. \ref{fig:B}(b) actually shows the existence of weakly nonlinear NPSW or PPSW according to $\beta_{1}$ lies either within the interval $0 \leq \beta_{1} < \beta_{1c}$ or within the interval $\beta_{3c} < \beta_{1} < \beta_{M}$. It is also important to note that for $0 \leq \beta_{1} < \beta_{1c}$ or $\beta_{3c} < \beta_{1} < \beta_{M}$, there does not exist any solitary structure at $M=M_{c}$.   

\begin{figure}
\begin{center}
  \includegraphics{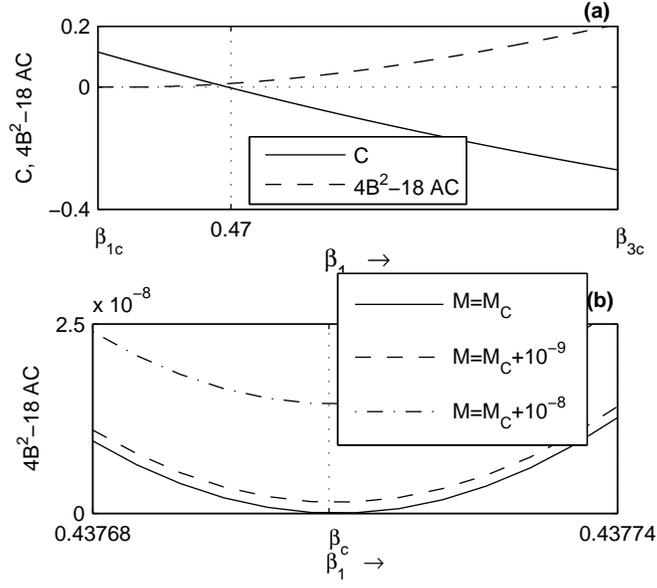}
  \caption{(a) $C$ and $4B^{2}-18AC$ is plotted against $\beta_{1}$ for $\beta_{1c}\leq \beta_{1} \leq \beta_{3c}$. (b) $4B^{2}-18AC$ is plotted against $\beta_{1}$ in a small neighborhood of $\beta_{c}$. \label{fig:C} }
\end{center}
\end{figure}

To recover the coexistence of small amplitude NPSWs and PPSWs along with PPDLs when $\beta_{1}$ lies within the interval $\beta_{1c} \leq \beta_{1} \leq \beta_{3c}$, we consider the following modified energy integral with $n=4$.
\begin{eqnarray}\label{energy int4}
\frac{1}{2} \bigg(\frac{d\phi}{d\xi}\bigg)^{2}+ V_{4} (\phi) = 0.
\end{eqnarray}
Here $V_{4}(\phi)$ is given by
\begin{eqnarray}\label{v4}
V_{4} (\phi) =\frac{1}{2} A \phi^{2} + \frac{1}{3} B \phi^{3} +
\frac{1}{4} C \phi^{4},
\end{eqnarray}
where
\begin{eqnarray}\label{C}
C = \frac{1}{6}\bigg[\frac{3 (5 M^{4} + 30 M^{2} \sigma_{d} + 9
\sigma_{d}^{2})}{(M^{2} - 3
\sigma_{d})^{5}}-\frac{1+3\beta_{1}+\alpha^{3} \mu}{1 - \mu}
\bigg],
\end{eqnarray}
$A$ and $B$ are given by Eqs. (\ref{A}) and (\ref{B}), respectively. For $C = 0$, $V_{4}(\phi)$ is same as $V_{3}(\phi)$, so, we consider the case for $C \neq 0$. For $C \neq 0$, we can write $V_{4}(\phi)$ as 
\begin{eqnarray}
V_{4}(\phi) = \frac{C}{4} \phi^{2}(\phi-\phi_{m1})(\phi-\phi_{m2}) , \label{mv4}
\end{eqnarray}
where
\begin{eqnarray}
\phi_{m1} =  -\frac{6A}{2B+\sqrt{4B^{2}-18AC}}, \label{phim1}
\end{eqnarray}
\begin{eqnarray}
\phi_{m2} =  -\frac{6A}{2B-\sqrt{4B^{2}-18AC}}, \label{phim2}
\end{eqnarray}
From (\ref{phim1}) and (\ref{phim2}), we note the following facts.\\
(i) $\phi_{m1}$ and $\phi_{m2}$ both are real and distinct if $4B^{2}-18AC >0$.\\
(ii) $\phi_{m1}\phi_{m2}=2A/C$.\\
(iii) $(\phi-\phi_{m1})(\phi-\phi_{m2})<0$ if and only if $\min\{\phi_{m1},\phi_{m2}\}<\phi<\max\{\phi_{m1},\phi_{m2}\}$.\\
(iv) $\phi_{m1} \rightarrow 0$ and $\phi_{m2} \rightarrow 0$ as $M \rightarrow M_{c}+0$.\\ 
The necessary conditions for the existence of solitary structures of the energy integral (\ref{energy int4}) are $V_{4}(0)=V'_{4}(0)$ and $V''_{4}(0)<0$. The conditions $V_{4}(0)=0$ and $V'_{4}(0)=0$ are trivially satisfied whereas the condition $V''_{4}(0)<0$ gives $A<0$. Here also the condition $A < 0$ gives $M>M_{c}$. Therefore, as $A<0$, from (ii), we see that $\phi_{m1}\phi_{m2}<0$ if $C>0$ and in this case, $V_{4}(\phi)<0$ for all $\min\{\phi_{m1},\phi_{m2}\}<\phi<0$ as well as $V_{4}(\phi)<0$ for all $0<\phi<\max\{\phi_{m1},\phi_{m2}\}$ and consequently, both NPSW and PPSW exist simultaneously. On the other hand, if $C<0$, one cannot get coexistence of weakly nonlinear NPSWs and PPSWs, since for $C<0$, both $\phi_{m1}$ and $\phi_{m2}$ are either positive or negative. Again as $\phi_{m1} \rightarrow 0$ and $\phi_{m2} \rightarrow 0$ as $M \rightarrow M_{c}+0$, $V_{4}(\phi)$ effectively define the coexistence of both NPSWs and PPSWs when $\beta_{1}$ lies within the interval $\beta_{1c} \leq \beta_{1} \leq \beta_{3c}$ provided that in this subinterval of $\beta_{1}$, $C>0$ and $4B^{2}-18AC >0$. In FIG. \ref{fig:C}(a), $C$ and $4B^{2}-18AC$ are plotted against $\beta_{1}$ ($\beta_{1c} \leq \beta_{1} \leq \beta_{3c}$) for $M = M_{c}+0.00001$. A quick look of this figure may suggest that $4B^{2}-18AC = 0$ at some point of $\beta_{1}$. To remove this illusion, we draw $4B^{2}-18AC$ against $\beta_{1}$ (in a small neighborhood of $\beta_{c}$) for three different values of $M$ in FIG. \ref{fig:C}(b). This figure clearly shows that $4B^{2}-18AC=0$ at $\beta_{1}=\beta_{c}$ only when $M=M_{c}$. However, we have already found that at $\beta_{1}=\beta_{c}$, $\phi=0$ is a position of stable equilibrium, and consequently there does not exist any solitary structure of the energy integral at $M=M_{c}$. Therefore, for $M>M_{c}$, $4B^{2}-18AC>0$ and consequently, these two figures actually show the coexistence of weakly nonlinear NPSWs and PPSWs when $\beta_{1}$ lies within the interval $\beta_{1c} \leq \beta_{1} <0.47(<\beta_{3c})$. So, $V_{4}(\phi)$ is unable to produce coexistence of both NPSWs and PPSWs for all $\beta_{1}$ within $\beta_{1c} \leq \beta_{1} \leq \beta_{3c}$, even when we take values of $M$ smaller than $M = M_{c}+0.00001$. Again, in this interval of $\beta_{1}$, i.e., $\beta_{1c} \leq \beta_{1} <0.47$, the amplitude of both NPSWs and PPSWs tends to zero as $M \rightarrow M_{c}+0$. But for $\beta_{1c} \leq \beta_{1} <\beta_{c}$, we get PPSWs of finite amplitude at $M=M_{c}$ and for $\beta_{c}<\beta_{1}<0.47$, we get NPSWs of finite amplitude at $M=M_{c}$. However, both PPSWs and NPSWs collapse at $\beta_{1} = \beta_{c}$, i.e., except for a very small neighborhood of $\beta_{1} = \beta_{c}$, PPSWs having finite amplitude exist at $M=M_{c}$ and NPSWs of finite amplitude exist at $M=M_{c}$. So, it is not possible to apply small amplitude method to study the weakly nonlinear solitary structures in a parameter regime where finite amplitude solitary structure occurs at $M=M_{c}$, because in any small amplitude method, amplitude of solitary wave always tends to zero as $M \rightarrow M_{c}+0$ which contradicts the occurrence of finite amplitude solitary wave at $M=M_{c}$. There may be another possibility that $V_{4}(\phi)$ is not the perfect match of $V(\phi)$ to study weakly nonlinear solitary structures in $\beta_{1c} \leq \beta_{1} \leq \beta_{3c}$. In any case, one can find inconsistency to study weakly nonlinear solitary structures for $\beta_{1c} \leq \beta_{1} \leq \beta_{3c}$. Therefore, our conclusion is that if there exists a region in the parameter space where finite amplitude solitary structures exist at $M=M_{c}$, it is not possible to study weakly nonlinear solitary structures by means of any small amplitude method in that portion of the parameter space.

\section{\label{sec:conclusion}Conclusion}
The paper deals with the existence of solitary wave and double layer at $M=M_{c}$. The work is relevant to some recent numerical observations. \cite{Verheest10b, Baluku10, Verheest10c, Baluku10a} However, the existence of solitary wave at $M=M_{c}$ is limited by means of these numerical observations only. In this paper, our motivation is to find the clarifications behind these numerical observations. The analytical theory for the existence of solitary wave and double layer at $M=M_{c}$ has been provided for the first time in the literature. Moreover, this analytical theory is purely based on the physical interpretation of energy integral as discussed in Sec. II. Without having the prior knowledge of physics of energy integral, it is not possible to give the proper mathematical clarification of the above mentioned numerical observations. Therefore, the theorems presented in this manuscript are based on the physical consequences of energy integral.

On investigating DAS waves in Sec.\ref{sec:verify}, we have found the existence of double layer solution at $M=M_{c}$. We have also discussed in our theoretical section, the conditions under which double layer solution occurs at $M=M_{c}$. This observation is not specific for this problem only. Actually, in any plasma environments, if the curve $M=M_{D}$ tends to intersect the curve $M=M_{c}$ at some point of the compositional parameter space or solution space, then we can always get a double layer solution at $M=M_{c}$, where $M=M_{D} (>M_{c})$ corresponds to a double layer solution. We have mentioned this result without proof. The existence of double layer solution at $M=M_{c}$ has been observed for the first time in literature and thus, this is quiet a new result.

From the investigation of solitary structures at $M=M_{c}$ of dust acoustic waves in Sec.\ref{sec:verify}, we find that the solitons and double layer are not two distinct nonlinear structures even when $M=M_{c}$. Actually, double layer solution, if exists, must be the limiting structure of at least one sequence of solitons of same polarity even when $M=M_{c}$. The following simple theorem of real analysis gives the actual logic behind the numerical observation that solitary waves end with a double layer, if exists.\\
\textit{If $\{x_{n}\}$ is a monotonically increasing sequence of real numbers, then $\{x_{n}\}$ converges to its \textit{lub}. Furthermore, if the sequence is bounded above, then \textit{lub}$\{x_{n}\}$ is finite.}\\ 
This theorem gives the actual logic behind the numerical observation of Baboolal \textsl{et al.} \cite{Baboolal90} Actually, Baboolal \textsl{et al.} \cite{Baboolal90} numerically observed that the double layer solution is the ultimate position of solitons, i.e., solitary waves end with a double layer, if exists. This numerical observation of Baboolal \textsl{et al.} \cite{Baboolal90} has been referred by several authors in different plasma environments. If we explain any fact with proper justification by means of mathematics, we can apply it in any specific problem. In the present paper, with the help of simple theorems, we have extended the physical interpretation for the existence of the solitary structures of the well-known energy integral even when the Mach number $M$ assumes its smallest possible value $M_{c}$. 

\begin{acknowledgments}
The authors are grateful to Prof. K. P. Das for his valuable suggestions to prepare this manuscript. One of the authors (Animesh Das) is thankful to State Government Departmental Fellowship Scheme for providing research support.
\end{acknowledgments}

\newpage

\end{document}